\documentclass[12pt]{article}
\usepackage{latexsym}
\usepackage{amsmath,amsfonts}
\usepackage{times}
\allowdisplaybreaks[4]
\catcode`\@=11

\newcount\hour
\newcount\minute
\newtoks\amorpm \hour=\time\divide\hour by 60\minute
=\time{\multiply\hour by 60 \global\advance\minute by-\hour}
\edef\standardtime{{\ifnum\hour<12 \global\amorpm={am}%
        \else\global\amorpm={pm}\advance\hour by-12 \fi
        \ifnum\hour=0 \hour=12 \fi
        \number\hour:\ifnum\minute<10
        0\fi\number\minute\the\amorpm}}
\edef\militarytime{\number\hour:\ifnum\minute<10 0\fi\number\minute}

\def\draftlabel#1{{\@bsphack\if@filesw {\let\thepage\relax
   \xdef\@gtempa{\write\@auxout{\string
      \newlabel{#1}{{\@currentlabel}{\thepage}}}}}\@gtempa
   \if@nobreak \ifvmode\nobreak\fi\fi\fi\@esphack}
        \gdef\@eqnlabel{#1}}
\def\@eqnlabel{}
\def\@vacuum{}
\def\marginnote#1{}
\def\draftmarginnote#1{\marginpar{\raggedright\scriptsize\tt#1}}
\def\draft{
        \pagestyle{plain}
        \overfullrule=2pt
        \oddsidemargin -.5truein
        \def\@oddhead{\sl \phantom{\today\quad\militarytime} \hfil
        \smash{\Large\sl DRAFT} \hfil \today\quad\militarytime}
        \let\@evenhead\@oddhead
        \let\label=\draftlabel
        \let\marginnote=\draftmarginnote
        \def\ps@empty{\let\@mkboth\@gobbletwo
        \def\@oddfoot{\hfil \smash{\Large\sl DRAFT} \hfil}
        \let\@evenfoot\@oddhead}
        \def\@eqnnum{(\theequation)\rlap{\kern\marginparsep\tt\@eqnlabel}%
        \global\let\@eqnlabel\@vacuum}  }

\newcommand{\rf}[1]{(\ref{#1})}
\renewcommand{\theequation}{\thesection.\arabic{equation}}
\renewcommand{\thefootnote}{\fnsymbol{footnote}}
\newcommand{\newsection}{   % Numeration of eqs. is automatic
\setcounter{equation}{0}\section}

\def\appendix#1{\addtocounter{section}{1}\setcounter{equation}{0}
\renewcommand{\thesection}{\Alph{section}}
\section*{Appendix \thesection\protect\indent \parbox[t]{11.15cm}{#1}}
\addcontentsline{toc}{section}{Appendix \thesection\ \ \ #1}}

\def\be{\begin{equation}}
\def\ee{\end{equation}}
\def\beq{\begin{eqnarray}}
\def\eeq{\end{eqnarray}}

\def\parline{\,\partial\kern -0.55em /\,\,}

\def\half{{\frac{1}{2}}}

\def\HH{{\cal H}}
\def\LL{{\cal L}}
\def\MM{{\cal M}}
\def\PP{{\cal P}}

\def\TT{{\cal T}}

\def\phik{|\phi\rangle}
\def\phibr{\langle\phi|}
\def\xik{|\xi\rangle}
\def\xibr{\langle\xi|}

\def\lambdak{|\lambda\rangle}

\def\Tk{|T\rangle}
\def\PPk{|\PP\rangle}

\def\smponetwo{{\scriptscriptstyle [1,2]}}

\def\oplussm{{\scriptscriptstyle \oplus}}
\def\ominussm{{\scriptscriptstyle \ominus}}

\def\oplussm{{\scriptscriptstyle \oplus}}
\def\ominussm{{\scriptscriptstyle \ominus}}

\def\m{{\rm m}}

\def\alpar{\alpha\partial}
\def\albpar{\bar\alpha\partial}

\def\Cb{\bar{C}}
\def\Gb{\bar{G}}

\def\eb{{\bar{e}}}

\def\alphab{{\bar{\alpha}}}

\def\alphabf{{\boldsymbol{\alpha}}}
\def\phibf{{\boldsymbol{\phi}}}
\def\xibf{{\boldsymbol{\xi}}}
\def\mubf{{\boldsymbol{\mu}}}
\def\Pibf{{\boldsymbol{\Pi}}}
\def\partialbf{{\boldsymbol{\partial}}}

\def\Cbf{{\bf C}}

\def\Ksf{{\sf K}}
\def\Lsf{{\sf L}}

\begin{document}

%\draft

\begin{flushright}
FIAN-TD-2011-13 \hspace{1.5cm}{}~\\
arXiv: 1112.0976 [hep-th]\quad{}~\\
Modified August 2012 \hspace{1cm}{}~
\end{flushright}

\vspace{1cm}

\begin{center}

{\Large \bf Extended hamiltonian action for arbitrary spin fields

\bigskip in flat and AdS spaces }

\vspace{2.5cm}

R.R. Metsaev\footnote{ E-mail: metsaev@lpi.ru }

\vspace{1cm}

{\it Department of Theoretical Physics, P.N. Lebedev Physical Institute, \\
Leninsky prospect 53,  Moscow 119991, Russia }

\vspace{3.5cm}

{\bf Abstract}

\end{center}

Totally symmetric arbitrary spin massless and massive free fields in
flat and AdS spaces and conformal fields in flat space are studied.
Extended gauge invariant hamiltonian action for such fields is
obtained. The action is constructed out of  phase space fields and
Lagrange multipliers which are free of algebraic constraints. Gauge
transformations of the phase space fields and Lagrange multipliers
are derived. Use of the Poincar\'e parametrization of AdS space
allows us to treat fields in flat space and AdS space on equal
footing.

\newpage
\renewcommand{\thefootnote}{\arabic{footnote}}
\setcounter{footnote}{0}

%%%%%%%%%%%%%%%%%%%%%%%%%%%%%%%%%%%%%%%%%%%%%%%%%%%%%%%%%%%%%%%%%%%%%%%
%%%%%%%%%%%%%%%%%%%%%%%%%%%%%%%%%%%%%%%%%%%%%%%%%%%%%%%%%%%%%%%%%%%%%%%
\section{\large Introduction}
%%%%%%%%%%%%%%%%%%%%%%%%%%%%%%%%%%%%%%%%%%%%%%%%%%%%%%%%%%%%%%%%%%%%%%%
%%%%%%%%%%%%%%%%%%%%%%%%%%%%%%%%%%%%%%%%%%%%%%%%%%%%%%%%%%%%%%%%%%%%%%%

In view of the aesthetic features of extended hamiltonian approach
to the relativistic field dynamics a interest in this approach was
periodically renewed (see e.g.
Refs.\cite{Dirac:1951zz,Fradkin:1977hw}). The extended hamiltonian
approach provides systematic and self-contained way to study many
aspects of relativistic field dynamics. Progress in understanding
higher-spin field dynamics \cite{Vasiliev:1990en} has lead to
intensive and in-depth study of various aspects of $AdS$ field
dynamics. Lagrangian formulation of higher-spin fields was developed
many years ago in Refs.\cite{Fronsdal:1978rb,Fronsdal:1978vb}. By
now many interesting approaches to $AdS$ fields are known in the
literature. However we note that the extended hamiltonian
formulation of massless and massive higher-spin fields in flat and
AdS spaces of arbitrary dimensions and conformal fields in flat
space of arbitrary dimensions
has not yet been worked out.%
\footnote{ Discussion of hamiltonian formulation of massless
fermionic fields in $4d$ flat and $AdS_4$ spaces may be found in
Refs.\cite{Aragone:1979hw,Vasiliev:1987hv} (for some discussion of
massless bosonic fields in $AdS_4$ see Ref.\cite{Vasiliev:1987hv}).
Discussion of hamiltonian formulation of massive spin 3/2 fermionic
field in $4d$ flat space may be found in Ref.\cite{Rindani:1988wf}.
Hamiltonian formulation of conformal spin-2 field in $4d$ space is
considered in Ref.\cite{Lee:1982cp}.}

The purpose of this paper is to develop gauge invariant hamiltonian
approach to totally symmetric arbitrary spin massless and massive
fields in flat and AdS spaces and conformal fields in flat space. In
this paper we deal with free bosonic fields. Our approach to the
extended hamiltonian field dynamics can be summarized as follows.

\noindent {\bf i}) We start with Lagrangian formulation of massless
and massive fields in flat and AdS spaces and conformal fields in
flat space and use representation for Lagrangian in terms of
modified de Donder divergence obtained in
Refs.\cite{Metsaev:2008fs}-\cite{Metsaev:2007rw}. We consider fields
in $d$ dimensional flat space and $d+1$ dimensional AdS space. We
use the Poincar\'e parametrization of $AdS_{d+1}$ space in which the
Lorentz algebra $so(d-1,1)$ symmetries are realized manifestly. We
use the double-traceless higher-spin fields of the Lorentz algebra
$so(d-1,1)$. It is the use of such double-traceless fields and the
Poincar\'e parametrization of AdS space that allows us to treat
massless and massive fields in flat and AdS spaces and conformal
fields in flat space on equal footing.

\noindent {\bf ii}) Our extended hamiltonian action is formulated in terms of
$so(d-1)$ algebra fields. All fields appearing in our extended hamiltonian
formulation are free of algebraic constraints. Field content entering our
extended hamiltonian action involves phase space fields and Lagrange
multipliers. Number of the Lagrangian multipliers and half of the phase space
fields is equal to the number of gauge fields appearing in the Lagrangian
formulation.

Our paper  is organized as follows.

In section \ref{sec-02}, we review the Lagrangian formulation of
massless and massive fields in flat and AdS spaces and conformal
fields in flat space. We discuss representation for Lagrangian in
terms of the modified de Donder divergence found in
Refs.\cite{Metsaev:2008fs}-\cite{Metsaev:2007rw}.  Also, we review
realization of gauge symmetries of the Lagrangian.

Sec. \ref{sec-03} is devoted to extended hamiltonian formulation of
massless and massive fields in flat and AdS spaces and conformal
fields in flat space. We start with description of field content
appearing in our approach. After this we present our result for
extended hamiltonian action and the corresponding gauge
transformations.

In Appendix, we summarize our conventions and the notation.

%%%%%%%%%%%%%%%%%%%%%%%%%%%%%%%%%%%%%%%%%%%%%%%%%%%%%%%%%%%%%%%%%%%%%%%
%%%%%%%%%%%%%%%%%%%%%%%%%%%%%%%%%%%%%%%%%%%%%%%%%%%%%%%%%%%%%%%%%%%%%%%
\newsection{\large Gauge invariant Lagrangian via modified de Donder divergence }
\label{sec-02}
%%%%%%%%%%%%%%%%%%%%%%%%%%%%%%%%%%%%%%%%%%%%%%%%%%%%%%%%%%%%%%%%%%%%%%%
%%%%%%%%%%%%%%%%%%%%%%%%%%%%%%%%%%%%%%%%%%%%%%%%%%%%%%%%%%%%%%%%%%%%%%%

In metric like approach, gauge invariant Lagrangian for free massless fields
in flat and $AdS_4$ spaces was obtained in
Refs.\cite{Fronsdal:1978rb,Fronsdal:1978vb}, while gauge invariant Lagrangian
for free massive fields in flat and $AdS_{d+1}$ spaces was found in
Ref.\cite{Zinoviev:2001dt}.%
\footnote{ For arbitrary $d$, various gauge invariant formulations of
massless fields in $AdS_{d+1}$ were discussed in
Refs.\cite{Lopatin:1987hz,Metsaev:1999ui,Buchbinder:2001bs}. In earlier
literature, study of arbitrary spin massive field in flat space via
dimensional reduction may be found in
Refs.\cite{Rindani:1985pi,Aragone:1988yx,Rindani:1988gb}. Discussion of
various dimensional reduction techniques in $AdS$ may be found in
Refs.\cite{Metsaev:2000qb,Biswas:2002nk,Artsukevich:2008vy}. In recent years,
higher-spin gauge fields have also been extensively studied in the framework
of BRST approach (see e.g.
Refs.\cite{Buchbinder:2005ua}-\cite{Grigoriev:2011gp}). Frame-like approach
to massive fields was developed in
Refs.\cite{Zinoviev:2008ze,Ponomarev:2010st}. In the framework of light-cone
gauge, the higher-spin AdS fields were studied in
Refs.\cite{Metsaev:1999ui,Metsaev:2002vr,Metsaev:2003cu,Metsaev:2004ee}.}
In Refs.\cite{Metsaev:2008fs}-\cite{Metsaev:2007rw}, we noticed that
use of modified de Donder divergence simplifies considerably the
structure of gauge invariant Lagrangian. Representation of the gauge
invariant Lagrangian for massive field in flat space in terms of
modified de Donder divergence was obtained in
Ref.\cite{Metsaev:2008fs}, while representation of the gauge
invariant Lagrangian for massless and massive fields in AdS space in
terms of the modified de Donder divergence was found in
Refs.\cite{Metsaev:2008ks,Metsaev:2009hp}. Representation of the
gauge invariant Lagrangian for conformal field in flat space in
terms of modified de Donder divergence was obtained in
Refs.\cite{Metsaev:2007fq,Metsaev:2007rw}. Because representation of
the gauge invariant Lagrangian via the modified de Donder divergence
turns out to be helpful for the derivation of extended hamiltonian
action we start with review of our results in
Refs.\cite{Metsaev:2008fs}--\cite{Metsaev:2007rw}. Before proceeding
to the review we note that we use the Cartesian parametrization of
Minkowski space and the Poincar\'e parametrization of $AdS_{d+1}$
space (for the notation, see Appendix),%
\footnote{ In our approach, only $so(d-1,1)$ symmetries are realized
manifestly. The $so(d,2)$ symmetries of fields in $AdS_{d+1}$ could
be realized manifestly by using ambient space approach (see e.g.
Refs.\cite{Bekaert:2009fg}-\cite{Fotopoulos:2006ci}.)}
\beq
&& ds^2  =  dx^a dx^a\,, \hspace{5.1cm} \hbox{ for flat space},
\\
&& ds^2 = \frac{1}{z^2} (dx^a dx^a + dzdz)\,,   \hspace{3cm} \hbox{ for AdS
space}.
\eeq
The use of such parametrizations allows us, among other thing, to treat
fields in flat and AdS spaces on equal footing. We now begin our review with
the discussion of field contents.

\noindent {\bf Field content for massless field in $R^{d-1,1}$}.
Massless spin-$s$ field in $d$-dimensional flat space can be
described by the rank-$s$ totally symmetric tensor field of the
Lorentz algebra $so(d-1,1)$, \cite{Fronsdal:1978rb},
\be \label{man02-26112011-01}
\phibf^{a_1\ldots a_s}\,,
\ee
subject to the double-tracelessness constraint, $\phibf^{aabba_5\ldots
a_{s}}=0$. To simplify the presentation of gauge invariant action we use
oscillators $\alpha^a$ and introduce the following ket-vector:
\be
\label{man02-26112011-02} |\phibf\rangle \equiv \frac{1}{s!} \alpha^{a_1} \ldots \alpha^{a_s}
\phibf^{a_1\ldots a_s} |0\rangle\,.
\ee

\noindent {\bf Field content for massive field in $R^{d-1,1}$}. As is well
known \cite{Zinoviev:2001dt}, spin-$s$ massive field in flat space can be
described by the following set of fields
\be
\label{man02-26112011-03} \phibf^{a_1\ldots a_{s'}}\,, \hspace{1cm} s'=0,1,\ldots,s.
\ee
Fields in \rf{man02-26112011-03} with $s'=0$, $s'=1$, and $s'\geq 2$ are the
respective scalar, vector, and rank-$s'$ totally symmetric fields of the
Lorentz algebra $so(d-1,1)$. Fields in \rf{man02-26112011-03} with $s'\geq 4$
are double-traceless, $\phibf^{aabba_5\ldots a_{s'}}=0$.
To streamline the presentation of gauge invariant action we use oscillators
$\alpha^a$, $\zeta$ and introduce the following ket-vector:
\be \label{man02-26112011-04}
|\phibf\rangle \equiv \sum_{s'=0}^s \frac{\zeta^{s-s'} \alpha^{a_1} \ldots
\alpha^{a_{s'}}}{s'!\sqrt{(s - s')!}} \, \phibf^{a_1\ldots a_{s'}} |0\rangle\,.
\ee

\noindent {\bf Field content for massless field in $AdS_{d+1}$}. To discuss
gauge invariant formulation of spin-$s$ massless field in $AdS_{d+1}$ we use
the following set of fields in Ref.\cite{Metsaev:2008ks}::
\be \label{man02-26112011-05}
\phibf^{a_1\ldots a_{s'}}\,, \hspace{1cm} s'=0,1,\ldots,s\,.
\ee
Fields in \rf{man02-26112011-05} with $s'=0$, $s'=1$, and $s'\geq 2$ are the
respective scalar, vector, and rank-$s'$ totally symmetric fields of the
Lorentz algebra $so(d-1,1)$. Fields in \rf{man02-26112011-05} with $s'\geq 4$ are
double-traceless, $\phibf^{aabba_5\ldots a_{s'}}=0$.%
\footnote{ In Ref.\cite{Fronsdal:1978vb}, the spin-$s$ massless field in
$AdS_{d+1}$ is described by rank-$s$ totally symmetric doubletraceless tensor
field of the Lorentz algebra $so(d,1)$. Note that $so(d-1,1)$ tensorial
components of the tensor field in Ref.\cite{Fronsdal:1978vb} are not
double-traceless. The tensor field in Ref.\cite{Fronsdal:1978vb} is related
to our fields \rf{man02-26112011-05} by invertible transformation. This
invertible transformation is described in Ref.\cite{Metsaev:2008ks}.}
To discuss gauge invariant Lagrangian in easy-to-use form we use oscillators
$\alpha^a$, $\alpha^z$ to collect fields \rf{man02-26112011-05} into the
ket-vector
\be \label{man02-26112011-06}  |\phibf\rangle \equiv \sum_{s'=0}^s
\frac{\alpha_z^{s-s'} \alpha^{a_1} \ldots \alpha^{a_{s'}}}{s'!\sqrt{(s -
s')!}} \, \phibf^{a_1\ldots a_{s'}} |0\rangle\,.
\ee

\noindent {\bf Field content for massive field in $AdS_{d+1}$}. To
discuss gauge invariant formulation of spin-$s$ massive field in
$AdS_{d+1}$ we use the following set of fields in
Ref.\cite{Metsaev:2009hp}:
\be \label{man02-26112011-07}
\phibf_n^{a_1\ldots a_{s'}}\,, \qquad s'=0,1,\ldots,s-1,s\,, \qquad
n \in [s-s']_2\,,
\ee
(for notation, see \rf{sumnot02} in Appendix). Fields in
\rf{man02-26112011-07} with $s'=0$, $s'=1$, and $s'\geq 2$ are the
respective scalar, vector, and rank-$s'$ totally symmetric fields of
the Lorentz algebra $so(d-1,1)$. Fields in \rf{man02-26112011-07}
with $s'\geq 4$
are double-traceless, $\phibf_n^{aabba_5\ldots a_{s'}}=0$.%
\footnote{ In Ref.\cite{Zinoviev:2001dt}, the spin-$s$ massive field
in $AdS_{d+1}$ is described by the set of fields involving totally
symmetric doubletraceless tensor fields of the Lorentz algebra
$so(d,1)$. Note that $so(d-1,1)$ tensorial components of the tensor
fields in Ref.\cite{Zinoviev:2001dt} are not double-traceless. The
fields in Ref.\cite{Zinoviev:2001dt} are related to our fields
\rf{man02-26112011-07} by invertible transformation. This invertible
transformation is described in Ref.\cite{Metsaev:2009hp}.}
To streamline the presentation we use oscillators $\alpha^a$,
$\alpha^z$, $\zeta$ and collect fields \rf{man02-26112011-07} into
the ket-vector defined by
\be \label{man02-26112011-08}
|\phibf\rangle = \sum_{s'=0}^s\,\,\sum_{n\in [s-s']_2}
\frac{\zeta_{\phantom{z}}^{\frac{s-s'+n}{2}}
\alpha_z^{\frac{s-s'-n}{2}}\alpha^{a_1}\ldots
\alpha^{a_{s'}}}{s'!\sqrt{(\frac{s-s'+n}{2})! (\frac{s-s'-n}{2})!}}
\, \phibf_n^{a_1\ldots a_{s'}} |0\rangle\,.
\ee

\noindent {\bf Field content for conformal field $R^{d-1,1}$}. To
discuss ordinary-derivative formulation of spin-$s$ conformal field in flat space
we use the following set of fields in
Ref.\cite{Metsaev:2007fq,Metsaev:2007rw}:
\be \label{man02-13082012-01}
\phibf_{k'}^{a_1\ldots a_{s'}}\,, \qquad s'=0,1,\ldots,s-1,s\,,
\qquad k' \in [k_{s'}]_2\,, \qquad k_{s'} \equiv s'
+\frac{d-6}{2}\,.
\ee
Fields in \rf{man02-13082012-01} with $s'=0$, $s'=1$, and $s'\geq 2$
are the respective scalar, vector, and rank-$s'$ totally symmetric
fields of the Lorentz algebra $so(d-1,1)$. Fields in
\rf{man02-13082012-01} with $s'\geq 4$ are double-traceless,
$\phibf_{k'}^{aabba_5\ldots a_{s'}}=0$. To simplify the presentation
we use the oscillators $\alpha^a$, $\zeta$, $\upsilon^\oplussm$,
$\upsilon^\ominussm$, and collect fields \rf{man02-13082012-01} into
ket-vector $\phik$ defined by
\be
\label{phikdef02}  |\phibf\rangle \equiv   \sum_{s'=0}^s \sum_{k'\in
[k_{s'}]_2} \frac{\zeta^{s-s'}
(\upsilon^\oplussm)^{^{\frac{k_{s'}+k'}{2}}}
(\upsilon^\ominussm)^{^{\frac{k_{s'} - k'}{2}}}}{s'!
\sqrt{(s-s')!}(\frac{k_{s'} + k'}{2})!}\alpha^{a_1} \ldots
\alpha^{a_{s'}} \, \phibf_{k'}^{a_1\ldots a_{s'}}|0\rangle\,.
\ee

\noindent {\bf Lagrangian}. Gauge invariant action for fields in
flat and AdS spaces is given by
\beq
&& \label{man02-26112011-09} S = \int d^dx\, \LL \,, \hspace{3.9cm}
\hbox{ for fields in } \ R^{d-1,1},
\nonumber\\[-10pt]
&&
\\[-10pt]
&& \label{man02-26112011-10} S = \int d^dx\, dz\,  \LL \,,
\hspace{3.4cm} \hbox{ for fields in} \ AdS_{d+1},
\nonumber
\eeq
where Lagrangian we found is given by
\beq
\label{man02-26112011-11}
&& \LL =  - \half \langle \partial^a \phibf|\mubf | \partial^a \phibf\rangle
-\half \langle \phibf| \mubf \MM^2| \phibf\rangle + \half \langle \bar\Cbf\phibf||
\bar\Cbf\phibf\rangle\,,
\\
\label{man02-26112011-12} &&\hspace{1.3cm}  \bar\Cbf \equiv
\bar\alphabf\partialbf - \half \alphabf\partialbf \bar\alphabf^2 -
\eb_1\Pibf^\smponetwo + \half e_1 \bar\alphabf^2\,,
\\
\label{man02-26112011-13} && \hspace{1.3cm} \Cbf \equiv \alphabf\partialbf -
\half \alphabf^2 \bar\alphabf\partialbf - e_1 \Pibf^\smponetwo + \half \eb_1
\alphabf^2\,,
\\
\label{man02-26112011-14} && \hspace{1.3cm} \mubf \equiv 1-
\frac{1}{4}\alphabf^2\bar\alphabf^2\,,\qquad \Pibf^\smponetwo \equiv 1
-\alphabf^2\frac{1}{2(2N_\alphabf +d)}\bar\alphabf^2\,,
\eeq
$\phibr\equiv (\phik)^\dagger$, $\langle \bar\Cbf\phibf| \equiv(|
\bar\Cbf\phibf\rangle)^\dagger$, $|\bar\Cbf\phibf\rangle \equiv \bar\Cbf|\phibf\rangle$,
and expressions like $\alphabf\partialbf$, $\alphabf^2$ are defined in Appendix (see \rf{manold-31102011-02}, \rf{manold-31102011-02a1}). We note that the Lagrangian for fields
in flat and AdS spaces and conformal fields is distinguished only by
the operators $\MM^2$, $e_1$ and $\eb_1$. To see this, we now
present the explicit form of these operators in turn.

\noindent {\bf Operators $\MM^2$, $e_1$, $\eb_1$ for massless field
in $R^{d-1,1}$}:
\be \label{man02-27112011-01}
\MM^2 = 0\,,\qquad e_1 = 0\,, \qquad  \eb_1 =  0\,.
\ee

\noindent {\bf Operators $\MM^2$, $e_1$, $\eb_1$ for massive field
in $R^{d-1,1}$}, \cite{Metsaev:2008fs}:
\beq
\label{man02-27112011-02} && \MM^2 = \m^2\,,
\qquad
e_1 = \m \zeta e_\zeta \,, \qquad \eb_1 =  - \m e_\zeta \bar\zeta \,,
\\
\label{man02-27112011-03} && e_\zeta \equiv
\Bigl(\frac{2s+d-4-N_\zeta}{2s+d-4-2N_\zeta}\Bigr)^{1/2}\,.
\eeq
In \rf{man02-27112011-02}, $\m$ stands for the commonly used mass
parameter of the massive field.

\noindent {\bf Operators $\MM^2$, $e_1$, $\eb_1$ for massless field
in $AdS_{d+1}$}, \cite{Metsaev:2008ks}:
\beq
&& \MM^2 = - \partial_z^2 + \frac{1}{z^2} (\nu^2-\frac{1}{4})\,,
\\[5pt]
&& - e_1 = \alpha^z e_z \TT_{\nu-\half} \,,
\qquad
- \eb_1 = \TT_{-\nu + \half} e_z \bar\alpha^z \,,
\\
&& \TT_\nu \equiv \partial_z+ \frac{\nu}{z}\,,
\qquad
\nu \equiv s+ \frac{d-4}{2} - N_z\,,
\\
\label{man02-27112011-07}  && e_z \equiv \Bigl(\frac{2s+ d-4 - N_z}{2s+ d-4 -
2N_z}\Bigr)^{1/2}\,.
\eeq

\noindent {\bf Operators $\MM^2$, $e_1$, $\eb_1$ for massive field
in $AdS_{d+1}$}, \cite{Metsaev:2009hp}:
\beq
&& \MM^2 \equiv - \partial_z^2 + \frac{1}{z^2} (\nu^2-\frac{1}{4})\,,
\\
&& - e_1 = \zeta r_\zeta \TT_{ -\nu - \half} + \alpha^z r_z \TT_{\nu-\half}
\,,
\qquad
- \eb_1 = \TT_{\nu + \half}  r_\zeta \bar\zeta  + \TT_{-\nu + \half} r_z
\bar\alpha^z \,,
\\
&& \TT_\nu \equiv \partial_z+ \frac{\nu}{z}\,,
\qquad
\nu \equiv \kappa + N_\zeta - N_z\,,
\\
\label{man02-27112011-08} && r_\zeta = \left(\frac{(s+\frac{d-4}{2}
-N_\zeta)(\kappa - s-\frac{d-4}{2} + N_\zeta)(\kappa + 1 +
N_\zeta)}{2(s+\frac{d-4}{2}-N_\zeta - N_z)(\kappa +N_\zeta -N_z) (\kappa+
N_\zeta - N_z +1)}\right)^{1/2}\,,
\\[7pt]
\label{man02-27112011-09}  && r_z = \left(\frac{(s+\frac{d-4}{2} -N_z)(\kappa
+ s + \frac{d-4}{2} - N_z)(\kappa - 1 -
N_z)}{2(s+\frac{d-4}{2}-N_\zeta-N_z)(\kappa + N_\zeta - N_z) (\kappa +N_\zeta
- N_z -1)}\right)^{1/2}\,,
\\
\label{man02-27112011-04}  && \kappa \equiv \sqrt{\m^2 + \Bigl( s+
\frac{d-4}{2}\Bigr)^2}\,.
\eeq
In \rf{man02-27112011-04}, $\m$ stands for the commonly used mass
parameter of the spin-$s$ massive field in $AdS_{d+1}$.

\noindent {\bf Operators $\MM^2$, $e_1$, $\eb_1$ for conformal field
in $R^{d-1,1}$}, \cite{Metsaev:2007rw}:
\be
\MM^2 = \upsilon^\ominussm \bar\upsilon^\ominussm \,,\qquad e_1 =
\zeta e_\zeta \bar\upsilon^\ominussm\,,\qquad \qquad \eb_1 = -
\upsilon^\ominussm e_\zeta \bar\zeta \,,
\ee
where $e_\zeta$ is given in \rf{man02-27112011-03}. The following
remarks are in order.

\noindent {\bf i}) It is the quantity $\bar\Cbf|\phibf\rangle$ that
we refer to as  modified de Donder divergence. Only for the case of
massless field in flat space this modified de Donder divergence
coincides with the standard de Donder divergence. From
\rf{man02-26112011-11},\rf{man02-26112011-12}, we see that many
complicated terms contributing to the Lagrangian are collected into
the modified de Donder divergence. Thus, as we have promised, use of
the modified de Donder divergence allows us to
simplify considerably a structure of the Lagrangian.%
\footnote{ Because our modified de Donder gauge leads to considerably
simplified analysis of AdS field dynamics we believe that this gauge might
also be useful for better understanding of various aspects of AdS/QCD
correspondence which are discussed e.g. in Ref.\cite{Gutsche:2011bu}.
Interesting applications of the {\it standard} de Donder-Feynman gauge to the
various problems of higher-spin fields may be found in
Refs.\cite{Guttenberg:2008qe,Manvelyan:2008ks,Fotopoulos:2009iw}.}

\noindent {\bf ii}) If we represent the $\langle
\bar\Cbf\phibf|\bar\Cbf\phibf\rangle$ contribution to the Lagrangian
in terms of the derivatives and oscillators, then, for the case of
massless and massive fields in flat space, Lagrangian given in
\rf{man02-26112011-11} takes the same form as in
Refs.\cite{Fronsdal:1978rb,Zinoviev:2001dt}. For the case of
massless and massive fields in $AdS_{d+1}$, in order to cast the
Lagrangians in Refs.\cite{Fronsdal:1978rb,Zinoviev:2001dt} into the
form given in \rf{man02-26112011-11} we use our set of the
$so(d-1,1)$ algebra double -traceless gauge fields. We recall that,
in Refs.\cite{Fronsdal:1978rb,Zinoviev:2001dt}, the Lagrangians of
massless and massive fields in $AdS_{d+1}$ are formulated in terms
of $so(d,1)$ algebra double-traceless gauge fields. Our gauge fields
are related to  gauge fields used in
Refs.\cite{Fronsdal:1978rb,Zinoviev:2001dt} by invertible
transformations. The invertible transformations are described in
Refs.\cite{Metsaev:2008ks,Metsaev:2009hp}.

\noindent {\bf iii}) Representation for Lagrangian in
\rf{man02-26112011-11} -\rf{man02-26112011-14} is universal and is
valid for arbitrary Poincar\'e invariant theory. Various Poincar\'e
invariant theories are distinguished by the operators $\MM^2$,
$e_1$, $\eb_1$. Namely, the dependence of the operators $\Cbf$,
$\bar\Cbf$ on the oscillators $\alpha^a$, $\bar\alpha^a$ and the
flat derivative $\partial^a$ takes the same form for massless and
massive fields in flat and $AdS$ spaces and conformal fields in flat
space. In other words, the operators $\Cbf$, $\bar\Cbf$ for massless
and massive fields in flat and AdS spaces and conformal fields in
flat space are distinguished only by the operators $e_1$ and
$\eb_1$.

\noindent {\bf iv}) Representation for Lagrangian given in
\rf{man02-26112011-11} turns out to be especially helpful for the
study of AdS/CFT duality for arbitrary spin massless and massive
bulk AdS fields and the corresponding boundary current and shadow
fields (see Refs.\cite{Metsaev:2009ym}-\cite{Metsaev:2011uy}.)

{\bf Gauge symmetries}. We now discuss gauge symmetries of Lagrangian given
in \rf{man02-26112011-11}. We begin with the description of gauge
transformation parameters involved in gauge transformations of gauge fields.
We discuss the gauge transformation parameters in turn.

\noindent {\bf Gauge transformations parameter for massless field in
$R^{d-1,1}$}. To discuss gauge symmetries of spin-$s$ massless field
in flat space we use the gauge transformation parameter
\cite{Fronsdal:1978rb},
\be \label{man02-26112011-15}
\xibf^{a_1\ldots a_{s-1}}\,,
\ee
which is rank-$(s-1)$ totally symmetric tensor field of the Lorentz algebra
$so(d-1,1)$. For $s\geq 3$ this parameter is traceless, $\xibf^{aaa_3\ldots
a_{s-1}} = 0$ (see Ref.\cite{Fronsdal:1978rb}). To simplify the presentation
we use the oscillator $\alpha^a$ and introduce the ket-vector
\be \label{man02-26112011-16}
|\xibf\rangle  \equiv \frac{1}{(s-1)!} \alpha^{a_1} \ldots \alpha^{a_{s-1}}
\xibf^{a_1\ldots a_{s-1}} |0\rangle\,.
\ee
\noindent {\bf Gauge transformations parameters for massive field in
$R^{d-1,1}$}. Gauge symmetries of spin-$s$ massive field in flat space are
described by the following set of gauge transformations parameters
in Ref.\cite{Zinoviev:2001dt}:
\be \label{man02-26112011-17}
\xibf^{a_1\ldots a_{s'}}\,, \hspace{1cm} s'=0,1,\ldots,s-1.
\ee
Gauge transformation parameters in \rf{man02-26112011-17} with $s'=0$,
$s'=1$, and $s'\geq 2$ are the respective scalar, vector, and rank-$s'$
totally symmetric fields of the Lorentz algebra $so(d-1,1)$. Gauge
transformation parameters in \rf{man02-26112011-17} with $s'\geq 2$ are
traceless, $\xibf^{aaa_3\ldots a_{s'}}=0$. To streamline
the presentation  we use the oscillators $\alpha^a$, $\zeta$ and introduce
the following ket-vector:
\be \label{man02-26112011-18}
|\xibf\rangle \equiv \sum_{s'=0}^{s-1} \frac{\zeta^{s-1-s'} \alpha^{a_1} \ldots
\alpha^{a_{s'}}}{s'!\sqrt{(s -1 - s')!}} \, \xibf^{a_1\ldots a_{s'}}
|0\rangle\,.
\ee

\noindent {\bf Gauge transformations parameters for massless field in
$AdS_{d+1}$}. To discuss gauge symmetries of spin-$s$ massless field in
$AdS_{d+1}$ we use the following set of gauge transformation parameters in
Ref.\cite{Metsaev:2008ks}:
\be
\label{man02-26112011-19} \xibf^{a_1\ldots a_{s'}}\,, \hspace{1cm}
s'=0,1,\ldots,s-1\,.
\ee
Gauge transformation parameters in \rf{man02-26112011-19} with $s'=0$,
$s'=1$, and $s'\geq 2$ are the respective scalar, vector, and rank-$s'$
totally symmetric fields of the Lorentz algebra $so(d-1,1)$. The gauge
transformation parameters in \rf{man02-26112011-19} with $s'\geq 2$
are traceless, $\xibf^{aaa_3\ldots a_{s'}}=0$.%
\footnote{ In Ref.\cite{Fronsdal:1978vb}, gauge symmetries of spin-$s$
massless field in $AdS_{d+1}$ are described by gauge transformation parameter
which is rank-$(s-1)$ totally symmetric traceless tensor field of the Lorentz
algebra $so(d,1)$. Note that $so(d-1,1)$ tensorial components of this gauge
transformation parameter are not traceless. Gauge transformation parameter in
Ref.\cite{Fronsdal:1978vb} is related to our gauge transformation parameters
\rf{man02-26112011-19} by invertible transformation described in
Ref.\cite{Metsaev:2008ks}.}
To simplify the presentation we use the oscillators $\alpha^a$, $\alpha^z$
and collect gauge transformation parameters \rf{man02-26112011-19} into the
ket-vector given by
\be \label{man02-26112011-20}
|\xibf\rangle \equiv \sum_{s'=0}^{s-1} \frac{\alpha_z^{s-1-s'} \alpha^{a_1} \ldots
\alpha^{a_{s'}}}{s'!\sqrt{(s-1 - s')!}} \, \xibf^{a_1\ldots a_{s'}}
|0\rangle\,.
\ee

\noindent {\bf Gauge transformations parameters for massive field in
$AdS_{d+1}$}.  To describe gauge symmetries of spin-$s$ massive
field in $AdS_{d+1}$ we use the following set of gauge
transformation parameters in Ref.\cite{Metsaev:2009hp}:
\be \label{man02-26112011-21}
\xibf_n^{a_1\ldots a_{s'}}\,, \qquad s'=0,1,\ldots, s-1\,, \qquad n
\in [s-1-s']_2\,.
\ee
Gauge transformation parameters in \rf{man02-26112011-21} with
$s'=0$, $s'=1$, and $s'\geq 2$ are the respective scalar, vector,
and rank-$s'$ totally symmetric fields of the Lorentz algebra
$so(d-1,1)$. The gauge transformation parameters in
\rf{man02-26112011-21} with $s'\geq 2$
are traceless, $\xibf_n^{aaa_3\ldots a_{s'}}=0$.%
\footnote{ In Ref.\cite{Zinoviev:2001dt}, gauge symmetries of
spin-$s$ massive field in $AdS_{d+1}$ are described by gauge
transformation parameters which are totally symmetric traceless
tensor fields of the Lorentz algebra $so(d,1)$. The $so(d-1,1)$
tensorial components of these gauge transformation parameters are
not traceless. Gauge transformation parameters in
Ref.\cite{Zinoviev:2001dt} are related to our gauge transformation
parameters \rf{man02-26112011-21} by invertible transformation
described in Ref.\cite{Metsaev:2009hp}.}
To streamline the presentation of gauge transformations we use
oscillators $\alpha^a$, $\alpha_z$, $\zeta$ and collect gauge
transformation parameters \rf{man02-26112011-21} into the ket-vector
defined by
\be \label{man02-26112011-22}
|\xibf\rangle = \sum_{s'=0}^{s-1}\,\,\sum_{n\in [s-1-s']_2}
\frac{\zeta_{\phantom{z}}^{\frac{s-1-s'+n}{2}}
\alpha_z^{\frac{s-1-s'-n}{2}}\alpha^{a_1}\ldots
\alpha^{a_{s'}}}{s'!\sqrt{(\frac{s-1-s'+n}{2})!
(\frac{s-1-s'-n}{2})!}} \, \xibf_n^{a_1\ldots a_{s'}} |0\rangle\,.
\ee

\noindent {\bf Gauge transformations parameters for conformal field
in $R^{d-1,1}$}.  To describe gauge symmetries of spin-$s$ conformal
field  we use the following set of gauge transformation parameters
in Ref.\cite{Metsaev:2007fq,Metsaev:2007rw}:
\be \label{13082012-07}
\xibf_{k'-1}^{a_1\ldots a_{s'}}\,,\hspace{1.5cm}
s'=0,1,\ldots,s-1\,,
\hspace{1.5cm} k' \in  [k_{s'}+1]_2\,.
\ee
Gauge transformation parameters in \rf{13082012-07} with $s'=0$,
$s'=1$, and $s'\geq 2$ are the respective scalar, vector, and
rank-$s'$ totally symmetric fields of the Lorentz algebra
$so(d-1,1)$. The gauge transformation parameters in \rf{13082012-07}
with $s'\geq 2$ are traceless, $\xibf_n^{aaa_3\ldots a_{s'}}=0$. To
simplify the presentation of we use oscillators $\alpha^a$, $\zeta$,
$\upsilon^\oplussm$, $\upsilon^\ominussm$ and collect gauge
transformation parameters \rf{13082012-07} into the ket-vector
defined by
\be
|\xibf\rangle \equiv \sum_{s'=0}^{s-1} \sum_{k'\in [k_{s'}+1]_2}
\frac{\zeta^{s-1-s'} (\upsilon^\oplussm)^{^{\frac{k_{s'}+1+k'}{2}}}
(\upsilon^\ominussm)^{^{\frac{k_{s'}+1-k'}{2}}}}{ s'!
\sqrt{(s-1-s')!} (\frac{k_{s'}+1+k'}{2})!}\alpha^{a_1} \ldots
\alpha^{a_{s'}} \, \xibf_{k'-1}^{a_1\ldots a_{s'}}
|0\rangle\,.\qquad
\ee

Having represented the field contents and gauge transformation
parameters in terms of the ket-vectors $|\phibf\rangle$ and
$|\xibf\rangle$ we note that the gauge transformations can entirely
be presented in terms of these ket-vectors. The representation for
gauge transformations found in
Refs.\cite{Metsaev:2008fs}-\cite{Metsaev:2007rw} is given by
\be \label{man02-26112011-23} \delta |\phibf\rangle = G |\xibf\rangle \,,
\qquad G \equiv \alphabf\partialbf - e_1 - \alphabf^2 \frac{1}{2N_\alphabf
+d-2}\eb_1 \,. \ee
For the cases of massless and massive fields in flat space, gauge
transformations \rf{man02-26112011-23} coincide with the respective
gauge transformations found in Refs.\cite{Fronsdal:1978rb} and
Ref.\cite{Zinoviev:2001dt}. For the case of massless and massive
fields in $AdS_{d+1}$, in order to cast the gauge transformations in
Refs.\cite{Fronsdal:1978vb,Zinoviev:2001dt} into the form given in
\rf{man02-26112011-23} we use our set of the $so(d-1,1)$ algebra
double-traceless gauge fields and the $so(d-1,1)$ algebra traceless
gauge transformation parameters.

%%%%%%%%%%%%%%%%%%%%%%%%%%%%%%%%%%%%%%%%%%%%%%%%%%%%%%%%%%%%%%%%%%%%%%%
%%%%%%%%%%%%%%%%%%%%%%%%%%%%%%%%%%%%%%%%%%%%%%%%%%%%%%%%%%%%%%%%%%%%%%%
\newsection{ \large Extended gauge invariant hamiltonian action }
\label{sec-03}
%%%%%%%%%%%%%%%%%%%%%%%%%%%%%%%%%%%%%%%%%%%%%%%%%%%%%%%%%%%%%%%%%%%%%%%
%%%%%%%%%%%%%%%%%%%%%%%%%%%%%%%%%%%%%%%%%%%%%%%%%%%%%%%%%%%%%%%%%%%%%%%

We now discuss the extended gauge invariant hamiltonian action for massless
and massive fields in flat and $AdS$ spaces and conformal fields in flat space.
We begin our discussion with the
description of field contents. We discuss the field contents in turn.

{\bf Field content for massless field in $R^{d-1,1}$}. To discuss hamiltonian
action for spin-$s$ massless field in flat space we introduce the following set of
fields:
\beq
\label{25112011-01} && \phi^{i_1\ldots i_s}\,, \qquad \ \ \ \PP^{i_1\ldots
i_s}\,,
\\
\label{25112011-02}&& \phi^{i_1\ldots i_{s-3}}\,, \qquad \PP^{i_1\ldots
i_{s-3}}\,,
\\
\label{25112011-03} && \lambda^{i_1\ldots i_{s-1}}\,, \qquad
\lambda^{i_1\ldots i_{s-2}}\,.
\eeq
Fields in \rf{25112011-01}-\rf{25112011-03} are totally symmetric
{\it traceful} tensor fields of the $so(d-1)$ algebra. Thus, we see
that all our fields in \rf{25112011-01}-\rf{25112011-03} are free
from any constraints, i.e. we deal with unconstrained fields. We
note that fields in \rf{25112011-01}, \rf{25112011-02} are phase
space variables, while fields in \rf{25112011-03} are Lagrange
multipliers. The fields $\phi^{i_1\ldots i_s}$, $\phi^{i_1\ldots
i_{s-3}}$ and Lagrange multipliers \rf{25112011-03} are related to
field in \rf{man02-26112011-01} by invertible transformation. To
simplify the presentation we use the oscillators $\alpha^i$ and
collect fields \rf{25112011-01}-\rf{25112011-03} into the following
ket-vectors:
\beq
\label{25112011-04} && |\phi_s\rangle \equiv \frac{1}{s!} \alpha^{i_1} \ldots
\alpha^{i_s} \phi^{i_1\ldots i_s} |0\rangle\,,
\nonumber\\[-10pt]
&&
\\[-10pt]
\label{25112011-05} && |\phi_{s-3}\rangle \equiv \frac{1}{(s-3)!}
\alpha^{i_1} \ldots \alpha^{i_{s-3}} \phi^{i_1\ldots i_{s-3}}
|0\rangle\,,\qquad
\nonumber\\
\label{25112011-06} && |\PP_s\rangle \equiv \frac{1}{s!} \alpha^{i_1} \ldots
\alpha^{i_s} \PP^{i_1\ldots i_s} |0\rangle\,,
\nonumber\\[-10pt]
&&
\\[-10pt]
\label{25112011-07} && |\PP_{s-3}\rangle \equiv \frac{1}{(s-3)!} \alpha^{i_1}
\ldots \alpha^{i_{s-3}} \PP^{i_1\ldots i_{s-3}} |0\rangle\,,\qquad
\nonumber\\
\label{25112011-08} && |\lambda_{s-1}\rangle \equiv \frac{1}{(s-1)!}
\alpha^{i_1} \ldots \alpha^{i_{s-1}} \lambda^{i_1\ldots i_{s-1}}
|0\rangle\,,\qquad
\nonumber\\[-10pt]
&&
\\[-10pt]
\label{25112011-09} && |\lambda_{s-2}\rangle \equiv \frac{1}{(s-2)!}
\alpha^{i_1} \ldots \alpha^{i_{s-2}} \lambda^{i_1\ldots i_{s-2}}
|0\rangle\,.\qquad
\nonumber
\eeq

\noindent {\bf Field content for massive field in $R^{d-1,1}$}.  To develop
hamiltonian approach to spin-$s$ massive field in flat space we introduce the
following set of fields:
\beq
&& \phi_s^{i_1\ldots i_{s'}}\,,\qquad \PP_s^{i_1\ldots
i_{s'}}\,, \hspace{1cm} s'=0,1,\ldots,s\,,
\nonumber\\[-10pt]
\label{25112011-10} &&
\\[-10pt]
&& \phi_{s-3}^{i_1\ldots i_{s'}}\,,\qquad
\PP_{s-3}^{i_1\ldots i_{s'}}\,, \hspace{1cm} s'=0,1,\ldots,s-3\,,
\nonumber\\
&& \lambda_{s-1}^{i_1\ldots i_{s'}}\,,\hspace{3.3cm}
s'=0,1,\ldots,s-1\,,
\nonumber\\[-10pt]
\label{25112011-13} &&
\\[-10pt]
&& \lambda_{s-2}^{i_1\ldots i_{s'}}\,,\hspace{3.3cm}
s'=0,1,\ldots,s-2\,.
\nonumber
\eeq
Fields in \rf{25112011-10},\rf{25112011-13} with $s'=0$, $s'=1$, and $s'\geq
2$ are the respective scalar, vector and totally symmetric rank-$s'$ {\it
traceful} tensor fields of the $so(d-1)$ algebra. To simplify the
presentation we use oscillators $\alpha^i$, $\zeta$ and collect fields
\rf{25112011-10},\rf{25112011-13} into the following ket-vectors:
\beq
\label{25112011-14} && |\phi_s\rangle \equiv \sum_{s'=0}^s \frac{\zeta^{s-s'}
\alpha^{i_1} \ldots \alpha^{i_{s'}}}{s'!\sqrt{(s - s')!}} \,
\phi_s^{i_1\ldots i_{s'}} |0\rangle\,,
\nonumber\\[-10pt]
&&
\\[-10pt]
\label{25112011-15} && |\phi_{s-3}\rangle \equiv \sum_{s'=0}^{s-3}
\frac{\zeta^{s-3-s'} \alpha^{i_1} \ldots
\alpha^{i_{s'}}}{s'!\sqrt{(s-3 - s')!}} \, \phi_{s-3}^{i_1\ldots
i_{s'}} |0\rangle\,,
\nonumber\\
\label{25112011-16} && |\PP_s\rangle \equiv \sum_{s'=0}^s \frac{\zeta^{s-s'}
\alpha^{i_1} \ldots \alpha^{i_{s'}}}{s'!\sqrt{(s - s')!}} \, \PP_s^{i_1\ldots
i_{s'}} |0\rangle\,,
\nonumber\\[-10pt]
&&
\\[-10pt]
\label{25112011-17} && |\PP_{s-3}\rangle \equiv \sum_{s'=0}^{s-3}
\frac{\zeta^{s-3-s'} \alpha^{i_1} \ldots
\alpha^{i_{s'}}}{s'!\sqrt{(s-3 - s')!}} \, \PP_{s-3}^{i_1\ldots
i_{s'}} |0\rangle\,,
\nonumber\\
\label{25112011-18} && |\lambda_{s-1}\rangle \equiv \sum_{s'=0}^{s-1}
\frac{\zeta^{s-1-s'} \alpha^{i_1} \ldots \alpha^{i_{s'}}}{s'!\sqrt{(s-1 -
s')!}} \, \lambda_{s-1}^{i_1\ldots i_{s'}} |0\rangle\,,
\nonumber\\[-10pt]
&&
\\[-10pt]
\label{25112011-19} && |\lambda_{s-2}\rangle \equiv \sum_{s'=0}^{s-2}
\frac{\zeta^{s-2-s'} \alpha^{i_1} \ldots
\alpha^{i_{s'}}}{s'!\sqrt{(s-2 - s')!}} \, \lambda_{s-2}^{i_1\ldots
i_{s'}} |0\rangle\,.
\nonumber
\eeq

\noindent {\bf Field content for massless field in $AdS_{d+1}$}. To discuss
hamiltonian action for spin-$s$ massless field in AdS  space we introduce the
following set of fields:
\beq
&& \phi_s^{i_1\ldots i_{s'}}\,,\qquad \PP_s^{i_1\ldots
i_{s'}}\,, \hspace{1cm} s'=0,1,\ldots,s\,,
\nonumber\\[-10pt]
\label{25112011-20} &&
\\[-10pt]
&& \phi_{s-3}^{i_1\ldots i_{s'}}\,,\qquad
\PP_{s-3}^{i_1\ldots i_{s'}}\,, \hspace{1cm} s'=0,1,\ldots,s-3\,,
\nonumber\\
&& \lambda_{s-1}^{i_1\ldots i_{s'}}\,,\hspace{3.3cm}
s'=0,1,\ldots,s-1\,,
\nonumber\\[-10pt]
\label{25112011-23} &&
\\[-10pt]
&& \lambda_{s-2}^{i_1\ldots i_{s'}}\,, \hspace{3.3cm}
s'=0,1,\ldots,s-2\,.
\nonumber
\eeq
We note that fields in \rf{25112011-20},\rf{25112011-23} with $s'=0$, $s'=1$,
and $s'\geq 2$ are the respective scalar, vector and totally symmetric
rank-$s'$ {\it traceful} tensor fields of the $so(d-1)$ algebra. To simplify
the presentation we use oscillators $\alpha^i$, $\alpha^z$ and collect fields
\rf{25112011-20},\rf{25112011-23} into the following ket-vectors:
\beq
\label{25112011-24} && |\phi_s\rangle \equiv \sum_{s'=0}^s
\frac{\alpha_z^{s-s'} \alpha^{i_1} \ldots \alpha^{i_{s'}}}{s'!\sqrt{(s -
s')!}} \, \phi_s^{i_1\ldots i_{s'}} |0\rangle\,,
\nonumber\\[-10pt]
&&
\\[-10pt]
\label{25112011-25} && |\phi_{s-3}\rangle \equiv \sum_{s'=0}^{s-3}
\frac{\alpha_z^{s-3-s'} \alpha^{i_1} \ldots
\alpha^{i_{s'}}}{s'!\sqrt{(s-3 - s')!}} \, \phi_{s-3}^{i_1\ldots
i_{s'}} |0\rangle\,,
\nonumber\\
\label{25112011-26} && |\PP_s\rangle \equiv \sum_{s'=0}^s
\frac{\alpha_z^{s-s'} \alpha^{i_1} \ldots \alpha^{i_{s'}}}{s'!\sqrt{(s -
s')!}} \, \PP_s^{i_1\ldots i_{s'}} |0\rangle\,,
\nonumber\\[-10pt]
&&
\\[-10pt]
\label{25112011-27} && |\PP_{s-3}\rangle \equiv \sum_{s'=0}^{s-3}
\frac{\alpha_z^{s-3-s'} \alpha^{i_1} \ldots
\alpha^{i_{s'}}}{s'!\sqrt{(s-3 - s')!}} \, \PP_{s-3}^{i_1\ldots
i_{s'}} |0\rangle\,,
\nonumber\\
\label{25112011-28} && |\lambda_{s-1}\rangle \equiv \sum_{s'=0}^{s-1}
\frac{\alpha_z^{s-1-s'} \alpha^{i_1} \ldots \alpha^{i_{s'}}}{s'!\sqrt{(s-1 -
s')!}} \, \lambda_{s-1}^{i_1\ldots i_{s'}} |0\rangle\,,
\nonumber\\[-10pt]
&&
\\[-10pt]
\label{25112011-29} && |\lambda_{s-2}\rangle \equiv \sum_{s'=0}^{s-2}
\frac{\alpha_z^{s-2-s'} \alpha^{i_1} \ldots
\alpha^{i_{s'}}}{s'!\sqrt{(s-2 - s')!}} \, \lambda_{s-2}^{i_1\ldots
i_{s'}} |0\rangle\,.
\nonumber
\eeq

\noindent {\bf Field content for massive field in $AdS_{d+1}$}. To
develop hamiltonian approach to spin-$s$ massive field in AdS space
we introduce the following set of fields:
\beq
&& \phi_{s,n}^{i_1\ldots i_{s'}}\,, \qquad
\PP_{s,n}^{i_1\ldots i_{s'}}\,, \qquad n \in
[s-s']_2\,,\hspace{1.7cm} s'=0,1,\ldots,s\,,
\nonumber\\[-10pt]
\label{25112011-30} &&
\\[-10pt]
&& \phi_{s-3,n}^{i_1\ldots i_{s'}}\,, \qquad
\PP_{s-3,n}^{i_1\ldots i_{s'}}\,, \qquad n \in [s-3-s']_2\,, \qquad
s'=0,1,\ldots,s-3\,,
\nonumber\\
&& \lambda_{s-1,n}^{i_1\ldots i_{s'}}\,,
\hspace{3.2cm} n \in [s-1-s']_2\,,\qquad s'=0,1,\ldots,s-1\,,
\nonumber\\[-10pt]
\label{25112011-33} &&
\\[-10pt]
&& \lambda_{s-2,n}^{i_1\ldots i_{s'}}\,,
\hspace{3.2cm} n \in [s-2-s']_2\,,\qquad s'=0,1,\ldots,s-2\,. \qquad
\nonumber
\eeq
Fields in \rf{25112011-30},\rf{25112011-33} with $s'=0$, $s'=1$, and
$s'\geq 2$ are the respective scalar, vector and totally symmetric
rank-$s'$ {\it traceful} tensor fields of the $so(d-1)$ algebra. To
simplify the presentation we use oscillators $\alpha^i$, $\alpha_z$,
$\zeta$ and collect fields \rf{25112011-30},\rf{25112011-33} into
the following ket-vectors:
\beq
\label{25112011-34}&& |\phi_s\rangle = \sum_{s'=0}^s\,\,\sum_{n\in
[s-s']_2} \frac{\zeta_{\phantom{z}}^{\frac{s-s'+n}{2}}
\alpha_z^{\frac{s-s'-n}{2}}\alpha^{i_1}\ldots
\alpha^{i_{s'}}}{s'!\sqrt{(\frac{s-s'+n}{2})! (\frac{s-s'-n}{2})!}}
\, \phi_{s,n}^{i_1\ldots i_{s'}} |0\rangle\,,
\nonumber\\[-10pt]
&&
\\[-10pt]
\label{25112011-35} && |\phi_{s-3}\rangle =
\sum_{s'=0}^{s-3}\,\,\sum_{n\in [s-3-s']_2}
\frac{\zeta_{\phantom{z}}^{\frac{s-3-s'+n}{2}}
\alpha_z^{\frac{s-3-s'-n}{2}}\alpha^{i_1}\ldots
\alpha^{i_{s'}}}{s'!\sqrt{(\frac{s-3-s'+n}{2})!
(\frac{s-3-s'-n}{2})!}} \, \phi_{s-3,n}^{i_1\ldots i_{s'}}
|0\rangle\,,
\nonumber\\
\label{25112011-36} && |\PP_s\rangle = \sum_{s'=0}^s\,\,\sum_{n\in
[s-s']_2} \frac{\zeta_{\phantom{z}}^{\frac{s-s'+n}{2}}
\alpha_z^{\frac{s-s'-n}{2}}\alpha^{i_1}\ldots
\alpha^{i_{s'}}}{s'!\sqrt{(\frac{s-s'+n}{2})! (\frac{s-s'-n}{2})!}}
\, \PP_{s,n}^{i_1\ldots i_{s'}} |0\rangle\,,
\nonumber\\[-10pt]
&&
\\[-10pt]
\label{25112011-37} && |\PP_{s-3}\rangle =
\sum_{s'=0}^{s-3}\,\,\sum_{n\in [s-3-s']_2}
\frac{\zeta_{\phantom{z}}^{\frac{s-3-s'+n}{2}}
\alpha_z^{\frac{s-3-s'-n}{2}}\alpha^{i_1}\ldots
\alpha^{i_{s'}}}{s'!\sqrt{(\frac{s-3-s'+n}{2})!
(\frac{s-3-s'-n}{2})!}} \, \PP_{s-3,n}^{i_1\ldots i_{s'}}
|0\rangle\,,
\nonumber\\
\label{25112011-38} && |\lambda_{s-1}\rangle =
\sum_{s'=0}^{s-1}\,\,\sum_{n\in [s-1-s']_2}
\frac{\zeta_{\phantom{z}}^{\frac{s-1-s'+n}{2}}
\alpha_z^{\frac{s-1-s'-n}{2}}\alpha^{i_1}\ldots
\alpha^{i_{s'}}}{s'!\sqrt{(\frac{s-1-s'+n}{2})!
(\frac{s-1-s'-n}{2})!}} \, \lambda_{s-1,n}^{i_1\ldots i_{s'}}
|0\rangle\,,
\nonumber\\[-10pt]
&&
\\[-10pt]
\label{25112011-39} && |\lambda_{s-2}\rangle =
\sum_{s'=0}^{s-2}\,\,\sum_{n\in [s-2-s']_2}
\frac{\zeta_{\phantom{z}}^{\frac{s-2-s'+n}{2}}
\alpha_z^{\frac{s-2-s'-n}{2}}\alpha^{i_1}\ldots
\alpha^{i_{s'}}}{s'!\sqrt{(\frac{s-2-s'+n}{2})!
(\frac{s-2-s'-n}{2})!}} \, \lambda_{s-2,n}^{i_1\ldots i_{s'}}
|0\rangle\,.
\nonumber
\eeq

\noindent {\bf Field content for conformal field in $R^{d-1,1}$}. To
develop hamiltonian approach to spin-$s$ conformal field in flat
space we introduce the following set of fields:
\beq
&& \phi_{s,k'}^{i_1\ldots i_{s'}}\,, \qquad
\PP_{s,k'}^{i_1\ldots i_{s'}}\,, \qquad k' \in
[k_{s'}]_2\,,\hspace{2.4cm} s'=0,1,\ldots,s\,,
\nonumber\\[-10pt]
\label{13082012-03} &&
\\[-10pt]
&& \phi_{s-3,k'}^{i_1\ldots i_{s'}}\,, \qquad
\PP_{s-3,k'}^{i_1\ldots i_{s'}}\,, \qquad k' \in [k_{s'}+3]_2\,,
\hspace{1.7cm} s'=0,1,\ldots,s-3\,,
\nonumber\\
&& \lambda_{s-1,k'}^{i_1\ldots i_{s'}}\,, \hspace{3.2cm} k' \in
[k_{s'}+1]_2\,,\hspace{1.7cm} s'=0,1,\ldots,s-1\,,
\nonumber\\[-10pt]
\label{13082012-06} &&
\\[-10pt]
&& \lambda_{s-2,k'}^{i_1\ldots i_{s'}}\,,
\hspace{3.2cm} k' \in [k_{s'}+2]_2\,,\hspace{1.7cm}
s'=0,1,\ldots,s-2\,. \qquad
\nonumber
\eeq
Fields in \rf{13082012-03},\rf{13082012-06} with $s'=0$, $s'=1$, and
$s'\geq 2$ are the respective scalar, vector and totally symmetric
rank-$s'$ {\it traceful} tensor fields of the $so(d-1)$ algebra. To
simplify the presentation we use oscillators $\alpha^i$, $\zeta$,
$\upsilon^\oplussm$, $\upsilon^\ominussm$ and collect fields
\rf{13082012-03},\rf{13082012-06} into the following ket-vectors:
\beq
&& |\phi_s\rangle \equiv   \sum_{s'=0}^s \sum_{k'\in [k_{s'}]_2}
\frac{\zeta^{s-s'} (\upsilon^\oplussm)^{^{\frac{k_{s'}+k'}{2}}}
(\upsilon^\ominussm)^{^{\frac{k_{s'} - k'}{2}}}}{s'!
\sqrt{(s-s')!}(\frac{k_{s'} + k'}{2})!}\alpha^{i_1} \ldots
\alpha^{i_{s'}} \, \phi_{s,k'}^{i_1\ldots i_{s'}}|0\rangle\,,
\nonumber\\[-10pt]
&&
\\[-10pt]
&& |\phi_{s-3}\rangle \equiv   \sum_{s'=0}^{s-3} \sum_{k'\in
[k_{s'}+3]_2} \frac{\zeta^{s-3-s'}
(\upsilon^\oplussm)^{^{\frac{k_{s'}+3+k'}{2}}}
(\upsilon^\ominussm)^{^{\frac{k_{s'} + 3 - k'}{2}}}}{s'!
\sqrt{(s-3-s')!}(\frac{k_{s'} + 3 + k'}{2})!}\alpha^{i_1} \ldots
\alpha^{i_{s'}} \, \phi_{s-3,k'}^{i_1\ldots i_{s'}}|0\rangle\,,
\nonumber\\
&& |\PP_s\rangle \equiv   \sum_{s'=0}^s \sum_{k'\in [k_{s'}]_2}
\frac{\zeta^{s-s'} (\upsilon^\oplussm)^{^{\frac{k_{s'}+k'}{2}}}
(\upsilon^\ominussm)^{^{\frac{k_{s'} - k'}{2}}}}{s'!
\sqrt{(s-s')!}(\frac{k_{s'} + k'}{2})!}\alpha^{i_1} \ldots
\alpha^{i_{s'}} \, \PP_{s,k'}^{i_1\ldots i_{s'}}|0\rangle\,,
\nonumber\\[-10pt]
&&
\\[-10pt]
&& |\PP_{s-3}\rangle \equiv   \sum_{s'=0}^{s-3} \sum_{k'\in
[k_{s'}+3]_2} \frac{\zeta^{s-3-s'}
(\upsilon^\oplussm)^{^{\frac{k_{s'}+3+k'}{2}}}
(\upsilon^\ominussm)^{^{\frac{k_{s'}+3 - k'}{2}}}}{s'!
\sqrt{(s-3-s')!}(\frac{k_{s'} + 3 + k'}{2})!}\alpha^{i_1} \ldots
\alpha^{i_{s'}} \, \PP_{s-3,k'}^{i_1\ldots i_{s'}}|0\rangle\,,
\nonumber\\
&& |\lambda_{s-1}\rangle \equiv   \sum_{s'=0}^{s-1} \sum_{k'\in
[k_{s'}+1]_2} \frac{\zeta^{s-1-s'}
(\upsilon^\oplussm)^{^{\frac{k_{s'} + 1 + k'}{2}}}
(\upsilon^\ominussm)^{^{\frac{k_{s'} + 1 - k'}{2}}}}{s'!
\sqrt{(s-1-s')!}(\frac{k_{s'} + 1 + k'}{2})!}\alpha^{a_1} \ldots
\alpha^{a_{s'}} \, \lambda_{s-1,k'}^{i_1\ldots i_{s'}}|0\rangle\,,
\nonumber\\[-10pt]
&&
\\[-10pt]
&& |\lambda_{s-2}\rangle \equiv   \sum_{s'=0}^{s-2} \sum_{k'\in
[k_{s'}+2]_2} \frac{\zeta^{s-2-s'}
(\upsilon^\oplussm)^{^{\frac{k_{s'} + 2 +k'}{2}}}
(\upsilon^\ominussm)^{^{\frac{k_{s'} + 2 - k'}{2}}}}{s'!
\sqrt{(s-2-s')!}(\frac{k_{s'} + 2 + k'}{2})!}\alpha^{i_1} \ldots
\alpha^{i_{s'}} \, \lambda_{s-2,k'}^{i_1\ldots i_{s'}}|0\rangle\,.
\nonumber
\eeq

To summarize, fields, which we use for discussing the extended hamiltonian
approach to massless and massive fields in flat and AdS spaces, can be
collected into the following ket-vectors:
\be \label{25112011-40}  |\phi_s\rangle\,,\qquad |\phi_{s-3}\rangle\,,\qquad
|\PP_s\rangle\,,\qquad |\PP_{s-3}\rangle\,,\qquad
|\lambda_{s-1}\rangle\,,\qquad |\lambda_{s-2}\rangle\,.\ee
We note that fields $|\phi_s\rangle$, $|\phi_{s-3}\rangle$, $|\PP_s\rangle$,
$|\PP_{s-3}\rangle$ are phase space variables, while the fields
$|\lambda_{s-1}\rangle$, $|\lambda_{s-2}\rangle$ are Lagrange multipliers. In
order to obtain the gauge invariant hamiltonian description in easy-to-use
form we collect fields \rf{25112011-40}  into 2 vectors given by
\be \label{25112011-41}
\phik = \left(
\begin{array}{l}
|\phi_s\rangle
\\[5pt]
|\phi_{s-3}\rangle
\end{array}\right)\,,
\\
\qquad\qquad
\PPk = \left(
\begin{array}{l}
|\PP_s\rangle
\\[5pt]
|\PP_{s-3}\rangle
\end{array}\right)\,,
\qquad
\lambdak = \left(
\begin{array}{l}
|\lambda_{s-1}\rangle
\\[5pt]
|\lambda_{s-2}\rangle
\end{array}\right)\,.
\ee

{\bf Extended hamiltonian action}. Extended gauge invariant
hamiltonian action for massless and massive fields in flat and $AdS$
spaces and conformal field in flat space takes the form
\beq
\label{25112011-43} && S = \int dt\, d^{d-1} x \, \LL \,,
\hspace{3.4cm} \hbox{ for fields in} \ R^{d-1,1}\,,
\nonumber\\[-10pt]
&&
\\[-10pt]
\label{25112011-44} && S = \int dt\,d^{d-1} x dz  \, \LL \,,
\hspace{3cm} \hbox{ for fields in } \ AdS_{d+1}\,,
\nonumber
\eeq
where $\LL$ is phase-space Lagrangian. The phase-space Lagrangian we
found is given by
\beq
\label{25112011-45} \LL & = & \langle \PP|\dot\phi\rangle - \half \langle \PP
|\Ksf^{-1} |\PP \rangle + \langle \PP|\Lsf|\phi\rangle + \LL^* + \langle \lambda
|T\rangle\,,
\\
\label{25112011-46} && \LL^* \equiv \half \langle \phi | E^*
|\phi\rangle \,,
\\
\label{25112011-46a1} && \Tk \equiv   \Gb_\phi \PPk -
\Gb_\PP\phik\,,
\eeq
where dot stands for derivative with respect to time $t$. Operators
constructed out of the spatial derivative $\partial^i$ and the
oscillators are given by
\beq
&& \Delta_\m  \equiv  \partial^i\partial^i - \MM^2\,,
\\
&& \Ksf  \equiv   \Ksf_0 \pi_+  +  \Ksf_3\pi_-\,,
\\
&& \Lsf  \equiv    \Ksf_0^{-1} \alpha^2 n_{44} C_{23} \sigma_+\,,
\\
&& E^*  \equiv  \left( n_{00} \Delta_\m - C_{10} n_{00} \Cb_{10}
\right)\pi_+
\nonumber\\
&& \qquad  +  \left( \Ksf_3 \Delta_\m - \Cb_{23} n_{44} C_{23} +
\Cb_{23} n_{44} \bar\alpha^2 \Ksf_0^{-1} \alpha^2 n_{44} C_{23}
\right)\pi_-\,,
\\
&& \Gb_\phi \equiv    \Gb_{01} \pi_+ +   G_{32}\pi_-\,,
\\
&& \Gb_\PP  \equiv  G_{31} \sigma_+ + \Gb_{02} \sigma_-\,,
\eeq
where $2\times 2$ matrices $\pi_\pm$, $\sigma_\pm$ and operators
$C_{mn}$, $G_{mn}$, $n_{00}$, $n_{44}$, $\Ksf_0$, $\Ksf_3$ are given
in Appendix. We note that the operators $n_{00}$, $n_{44}$,
$\Ksf_0$, $\Ksf_3$ depend only on the spatial oscillators and are
independent of the spatial derivative. From \rf{25112011-45}, we see
that the fields $\phik$ and $\PPk$ are realized as phase space
variables, while the field $\lambdak$ is realized as Lagrange
multiplier.

{\bf Gauge transformations}. We now discuss realization of gauge
symmetries in the framework of extended hamiltonian gauge invariant
approach. We begin our discussion with the description of gauge
transformation parameters to be used for description of gauge
transformations. We discuss the gauge transformation parameters in
turn.

\noindent {\bf Gauge transformations parameters for massless field in
$R^{d-1,1}$}. In the framework of our hamiltonian gauge invariant approach,
gauge symmetries of spin-$s$ massless field in flat space are described by
the following two gauge transformation parameters:
\be
\label{25112011-48} \xi^{i_1\ldots i_{s-1}}\,, \qquad \xi^{i_1\ldots
i_{s-2}}\,.
\ee
For the corresponding values of $s$, the gauge transformation parameters in
\rf{25112011-48} are scalar, vector and {\it traceful} tensor fields of the
$so(d-1)$ algebra. We use the oscillators $\alpha^i$ to collect the
parameters into two ket-vectors given by
\beq
\label{25112011-49} && |\xi_{s-1}\rangle  \equiv \frac{1}{(s-1)!}
\alpha^{i_1} \ldots \alpha^{i_{s-1}} \xi^{i_1\ldots i_{s-1}} |0\rangle\,,
\nonumber\\[-10pt]
&&
\\[-10pt]
\label{25112011-50} && |\xi_{s-2}\rangle   \equiv \frac{1}{(s-2)!}
\alpha^{i_1} \ldots \alpha^{i_{s-2}} \xi^{i_1\ldots i_{s-2}}
|0\rangle\,.
\nonumber
\eeq

\noindent {\bf Gauge transformations parameters for massive field in
$R^{d-1,1}$}. In the framework of our hamiltonian gauge invariant approach,
gauge symmetries of spin-$s$ massive field in flat space are described by the
following set of gauge transformation parameters:
\beq
\label{25112011-51} && \xi_{s-1}^{i_1\ldots i_{s'}}\,, \hspace{1cm}
s'=0,1,\ldots,s-1\,,
\nonumber\\[-10pt]
&&
\\[-10pt]
\label{25112011-52}&& \xi_{s-2}^{i_1\ldots i_{s'}}\,, \hspace{1cm}
s'=0,1,\ldots,s-2\,.
\nonumber
\eeq
We note that gauge transformation parameters in \rf{25112011-51}
with $s'=0$, $s'=1$, and $s'\geq 2$ are the respective scalar,
vector, and rank-$s'$ {\it traceful} tensor fields of the $so(d-1)$
algebra. We use the oscillators $\alpha^i$, $\zeta$ to collect the
gauge transformation parameters into two ket-vectors given by
\beq
\label{25112011-53}&& |\xi_{s-1}\rangle \equiv \sum_{s'=0}^{s-1}
\frac{\zeta^{s-1-s'} \alpha^{i_1} \ldots \alpha^{i_{s'}}}{s'!\sqrt{(s -1 -
s')!}} \, \xi_{s-1}^{i_1\ldots i_{s'}} |0\rangle\,,
\nonumber\\[-10pt]
&&
\\[-10pt]
\label{25112011-54}&& |\xi_{s-2}\rangle \equiv \sum_{s'=0}^{s-2}
\frac{\zeta^{s-2-s'} \alpha^{i_1} \ldots
\alpha^{i_{s'}}}{s'!\sqrt{(s -2 - s')!}} \, \xi_{s-2}^{i_1\ldots
i_{s'}} |0\rangle\,.
\nonumber
\eeq

\noindent {\bf Gauge transformations parameters massless field in
$AdS_{d+1}$}. To discuss gauge symmetries of spin-$s$ massless AdS field in
the framework of hamiltonian gauge invariant approach, we use the following
set of gauge transformation parameters:
\beq
\label{25112011-55}&& \xi_{s-1}^{i_1\ldots i_{s'}}\,, \hspace{2cm}
s'=0,1,\ldots,s-1\,,
\nonumber\\[-10pt]
&&
\\[-10pt]
\label{25112011-56}&& \xi_{s-2}^{i_1\ldots i_{s'}}\,, \hspace{2cm}
s'=0,1,\ldots,s-2\,.
\nonumber
\eeq
Gauge transformation parameters in \rf{25112011-55} with $s'=0$,
$s'=1$, and $s'\geq 2$ are the respective scalar, vector, and
rank-$s'$ {\it traceful} tensor fields of the $so(d-1)$ algebra. We
use the oscillators $\alpha^i$, $\alpha_z$ to collect the gauge
transformation parameters into two ket-vectors given by
\beq
\label{25112011-57}&& |\xi_{s-1}\rangle \equiv \sum_{s'=0}^{s-1}
\frac{\alpha_z^{s-1-s'} \alpha^{i_1} \ldots \alpha^{i_{s'}}}{s'!\sqrt{(s -1 -
s')!}} \, \xi_{s-1}^{i_1\ldots i_{s'}} |0\rangle\,,
\nonumber\\[-10pt]
&&
\\[-10pt]
\label{25112011-58}&& |\xi_{s-2}\rangle \equiv \sum_{s'=0}^{s-2}
\frac{\alpha_z^{s-2-s'} \alpha^{i_1} \ldots
\alpha^{i_{s'}}}{s'!\sqrt{(s -2 - s')!}} \, \xi_{s-2}^{i_1\ldots
i_{s'}} |0\rangle\,.
\nonumber
\eeq

\noindent {\bf Gauge transformations parameters for massive field in
$AdS_{d+1}$}.  To discuss gauge symmetries of spin-$s$ massive AdS
field in the framework of hamiltonian gauge invariant approach, we
use the following set of gauge transformation parameters:
\beq
\label{25112011-59} && \xi_{s-1,n}^{i_1\ldots i_{s'}}\,, \qquad n
\in [s-1-s']_2\,,\qquad s'=0,1,\ldots, s-1\,,
\nonumber\\[-10pt]
&&
\\[-10pt]
\label{25112011-60}&& \xi_{s-2,n}^{i_1\ldots i_{s'}}\,, \qquad n \in
[s-2-s']_2\,,\qquad s'=0,1,\ldots, s-2\,.
\nonumber
\eeq
Gauge transformation parameters in \rf{25112011-59} with $s'=0$,
$s'=1$, and $s'\geq 2$ are the respective scalar, vector, and
rank-$s'$ {\it traceful} tensor fields of the $so(d-1)$ algebra. We
use the oscillators $\alpha^i$, $\alpha^z$, $\zeta$ to collect the
gauge transformation parameters into two ket-vectors given by
\beq
\label{25112011-61}&& |\xi_{s-1}\rangle =
\sum_{s'=0}^{s-1}\,\,\sum_{n\in [s-1-s']_2}
\frac{\zeta_{\phantom{z}}^{\frac{s-1-s'+n}{2}}
\alpha_z^{\frac{s-1-s'-n}{2}}\alpha^{i_1}\ldots
\alpha^{i_{s'}}}{s'!\sqrt{(\frac{s-1-s'+n}{2})!
(\frac{s-1-s'-n}{2})!}} \, \xi_{s-1,n}^{i_1\ldots i_{s'}}
|0\rangle\,,
\\
\label{25112011-62}&& |\xi_{s-2}\rangle =
\sum_{s'=0}^{s-2}\,\,\sum_{n\in [s-2-s']_2}
\frac{\zeta_{\phantom{z}}^{\frac{s-2-s'+n}{2}}
\alpha_z^{\frac{s-2-s'-n}{2}}\alpha^{i_1}\ldots
\alpha^{i_{s'}}}{s'!\sqrt{(\frac{s-2-s'+n}{2})!
(\frac{s-2-s'-n}{2})!}} \, \xi_{s-2,n}^{i_1\ldots i_{s'}}
|0\rangle\,.
\eeq

\noindent {\bf Gauge transformations parameters for conformal field
in $R^{d-1,1}$}.  To discuss gauge symmetries of spin-$s$ conformal
field in the framework of hamiltonian gauge invariant approach, we
use the following set of gauge transformation parameters:
\beq
\label{12082012-08} && \xi_{s-1,k'-1}^{i_1\ldots
i_{s'}}\,,\hspace{1.5cm} s'=0,1,\ldots,s-1\,,
\hspace{1.5cm} k' \in  [k_{s'}+1]_2\,.
\nonumber\\[-10pt]
&&
\\[-10pt]
\label{12082012-09} && \xi_{s-2,k'-1}^{i_1\ldots
i_{s'}}\,,\hspace{1.5cm} s'=0,1,\ldots,s-2\,,
\hspace{1.5cm} k' \in  [k_{s'}+2]_2\,.
\nonumber
\eeq
Gauge transformation parameters in \rf{12082012-08} with $s'=0$,
$s'=1$, and $s'\geq 2$ are the respective scalar, vector, and
rank-$s'$ {\it traceful} tensor fields of the $so(d-1)$ algebra. We
use the oscillators $\alpha^i$, $\zeta$, $\upsilon^\oplussm$,
$\upsilon^\ominussm$ to collect the parameters into two ket-vectors
given by
\beq
&& |\xi_{s-1}\rangle \equiv \sum_{s'=0}^{s-1} \sum_{k'\in
[k_{s'}+1]_2} \frac{\zeta^{s-1-s'}
(\upsilon^\oplussm)^{^{\frac{k_{s'}+1+k'}{2}}}
(\upsilon^\ominussm)^{^{\frac{k_{s'}+1-k'}{2}}}}{ s'!
\sqrt{(s-1-s')!} (\frac{k_{s'}+1+k'}{2})!}\alpha^{i_1} \ldots
\alpha^{i_{s'}} \, \xi_{s-1,k'-1}^{i_1\ldots i_{s'}}
|0\rangle\,,\qquad
\nonumber\\[-10pt]
&&
\\[-10pt]
&& |\xi_{s-2}\rangle \equiv \sum_{s'=0}^{s-2} \sum_{k'\in
[k_{s'}+2]_2} \frac{\zeta^{s-2-s'}
(\upsilon^\oplussm)^{^{\frac{k_{s'}+2+k'}{2}}}
(\upsilon^\ominussm)^{^{\frac{k_{s'}+2-k'}{2}}}}{ s'!
\sqrt{(s-2-s')!} (\frac{k_{s'}+2+k'}{2})!}\alpha^{i_1} \ldots
\alpha^{i_{s'}} \, \xi_{s-2,k'-1}^{i_1\ldots i_{s'}}
|0\rangle\,.\qquad
\nonumber
\eeq

To summarize, we note that, in all cases above-considered, the gauge
transformation parameters we are going to use for the description of gauge
symmetries of massless and massive fields in flat and AdS spaces can be
collected into two ket-vectors
\be \label{25112011-63} |\xi_{s-1}\rangle\,,\qquad \qquad
|\xi_{s-2}\rangle\,.\ee
As before, in order to obtain the gauge transformations in easy-to-use form
we collect gauge transformation parameters \rf{25112011-63} into 2 vector
given by

\be \label{25112011-64}
\xik = \left(
\begin{array}{l}
|\xi_{s-1}\rangle
\\[6pt]
|\xi_{s-2}\rangle
\end{array}\right)\,.
\ee

We now discuss gauge transformations. The gauge transformations can
entirely be presented in terms of ket-vectors above discussed. The
gauge transformations we found take the form
\beq
\label{man02-30112011-01} && \hspace{-1cm} \delta \phik =G_\phi \xik\,,
\\
\label{man02-30112011-02} && \hspace{-1cm} \delta \PPk =G_\PP \xik\,,
\\
\label{man02-30112011-03} &&  \hspace{-1cm} \delta \lambdak = |\dot\xi\rangle
+ G_\lambda \xik\,,
\\
&& G_\phi \equiv  G_{01} \pi_+ + \Gb_{32}\pi_-\,,
\\
&& G_\PP \equiv  G_{02} \sigma_+  + \Gb_{31} \sigma_-\,,
\\
&& G_\lambda \equiv   G_{12} \sigma_+ + \Gb_{21}\sigma_-\,,
\eeq
where explicit form of oparators $G_{mn}$ is given in Appendix. The
following remarks are in order.

\noindent {\bf i}) From \rf{man02-30112011-03} we see that gauge
transformations of the Lagrange multiplier $\lambdak$ involve time derivative
of the gauge transformation parameter, as it should be in extended
hamiltonian approach (see e.g. Ref.\cite{Fradkin:1977hw}).

\noindent {\bf ii}) Introducing Hamiltonian $H$ and gauge transformation
generating function $T_\xi$,
\beq
\label{mna02-01122011-03} &&  H \equiv \int d^{d-1}x\, \HH \,, \hspace{2.6cm}
\hbox{ for fields in } \ R^{d-1,1}\,,
\nonumber\\[-10pt]
&&
\\[-10pt]
\label{mna02-01122011-04} &&  H \equiv \int d^{d-1}x\, dz\, \HH \,,
\hspace{2.1cm} \hbox{ for fields in } \  AdS_{d+1}\,,
\nonumber
\eeq

\be \label{mna02-01122011-05} - \HH \equiv - \half \langle \PP |\Ksf^{-1}
|\PP \rangle + \langle \PP|\Lsf|\phi\rangle + \LL^* \,, \ee

\beq
&&  T_\xi \equiv \int d^{d-1}x\,\xibr \Tk \,, \hspace{2.6cm} \hbox{ for
fields in } \ R^{d-1,1}\,,
\nonumber\\[-10pt]
&&
\\[-10pt]
&& T_\xi \equiv \int d^{d-1}x\, dz\, \xibr \Tk \,, \hspace{2.1cm} \hbox{ for
fields in } \ AdS_{d+1}\,,
\nonumber
\eeq
where $\Tk$ is given in \rf{25112011-46a1}, we find that under gauge
transformations \rf{man02-30112011-01}-\rf{man02-30112011-03} the constraint
$\Tk$ and Hamiltonian $H$ transform as
\beq
\label{man02-01122011-01} && \delta \Tk =0 \,,
\\
\label{man02-01122011-02} && \delta H = T_{G_\lambda \xi}^{\vphantom{7pt}} \,.
\eeq
Relation \rf{man02-01122011-01} tells that the constraint $\Tk$ is invariant
under the gauge transformations, while from relation \rf{man02-01122011-02}
we learn that gauge variation of the Hamiltonian $H$ is proportional to the
constraint $\Tk$. In other words, the $\Tk$ is the first-class constraint.

\noindent {\bf iii}) Lagrangian \rf{25112011-45} implies the standard
equal-time Poisson bracket,
\beq
\label{mna02-01122011-06} && {} [\PPk,\phibr] = |\rangle \langle |
\delta^{d-1}(x-x')  \,, \hspace{3cm} \hbox{ for fields in } \ R^{d-1,1}\,,
\nonumber\\[-10pt]
&&
\\[-10pt]
\label{mna02-01122011-07} && {} [\PPk,\phibr] = |\rangle \langle |
\delta^{d-1}(x-x')\delta(z-z')  \,, \hspace{1.4cm} \hbox{ for fields
in } \ AdS_{d+1}\,,\qquad
\nonumber
\eeq
where $|\rangle\langle|$ stands for the unit operator on space of ket-vectors
given in \rf{25112011-41}. Using the Poisson bracket, we check that gauge
transformations of phase space variables $\phik$ and $\PPk$ given in
\rf{man02-30112011-01},\rf{man02-30112011-02} can be represented as
\beq
&& \label{23102011-10} \delta \phik = [\phik,T_\xi]\,,
\qquad\quad \delta \PPk = [\PPk,T_\xi]\,,
\eeq
as it should be in the framework of the extended hamiltonian approach. Also,
in terms of the Poisson bracket, gauge transformations given in
\rf{man02-01122011-01},\rf{man02-01122011-02} can be represented as
\beq
&& \label{23102011-14a1} [T_{\xi_1},T_{\xi_2}] = 0\,,
\\
&& \label{23102011-14a2} [T_\xi ,H] = -T_{G_\lambda\xi}^{\vphantom{7pt}} \,.
\eeq

{\bf iv}) As illustration, let us count physical D.o.F for spin-$s$ massless
field in $d$-dimensional flat space by using the extended hamiltonian
approach. Using notation $N_{s,n}$ for the dimension of the totally symmetric
rank-$s$ traceful tensor field of $so(n)$ algebra,
\be N_{s,n} = \frac{(s+n-1)!}{(n-1)!s!}\,,
\ee
we note that the dimensions of the fields $\phi^{i_1\ldots i_s}$ and
$\phi^{i_1\ldots i_{s-3}}$ are given by $N_{s,d-1}$ and $N_{s-3,d-1}$
respectively, while the dimensions of the Lagrange multipliers
$\lambda^{i_1\ldots i_{s-1}}$ and $\lambda^{i_1\ldots i_{s-2}}$ are given by
$N_{s-1,d-1}$ and $N_{s-2,d-1}$ respectively. Applying standard formula for
counting physical D.o.F (see e.g. Ref.\cite{Fradkin:1977hw}), we find the relation
\be \label{man02-04122011-01} N_{s,d-1} + N_{s-3,d-1} - N_{s-1,d-1} -
N_{s-2,d-1} = (2s+d-4)\frac{(s+d-5)!}{(d-4)!s!} \,. \ee
Number in r.h.s. in \rf{man02-04122011-01} is a dimension of totally
symmetric rank-$s$ traceless tensor field of $so(d-2)$ algebra. This
dimension is the number of physical D.o.F . for spin-$s$ massless field in
$d$-dimensional space-time.

To summarize, staring with the Lagrangian formulation of double-traceless higher-spin fields
we obtained the extended hamiltonian
action in terms of fields which are free of algebraic constraints. We believe
that the appearance of unconstrained fields in the extended hamiltonian
approach should streamline application of our
approach to the study of various aspects of higher-spin fields.%
\footnote{ As a side remark we note that at Lagrangian level many interesting
formulations in terms of unconstrained fields were developed in last years.
This is to say various Lagrangian formulations of higher-spin field dynamics
in terms of unconstrained fields are discussed in
Refs.\cite{Francia:2002aa}-\cite{Campoleoni:2008jq}.}
Also, we think that the power of hamiltonian methods will provide new
possibilities for analyzing equations of motion of AdS fields and studying
AdS/CFT correspondence.%
\footnote{ Discussion of interesting methods for analyzing  equations of
motion of fields in AdS space may be found in
Refs.\cite{Bolotin:1999fa}-\cite{Didenko:2011ir}.}

In conclusion, we note a number of the potentially interesting
generalizations and applications of our approach. This is to say that
although many methods for building interaction vertices for higher-spin
fields are known in the literature (see e.g.
Refs.\cite{Fradkin:1987ks}-\cite{Joung:2011ww}), constructing interaction
vertices for concrete field theoretical models of higher-spin fields is still
a challenging problem. We believe that use of the extended hamiltonian
approach will provide new interesting possibilities for studying this
important problem. Also we think that the extended hamiltonian approach we
discussed in this paper might be useful for the study of string theory in AdS
background \cite{Metsaev:1998it}-\cite{Metsaev:2000mv} and various aspects of
AdS/CFT correspondence along the lines in
Refs.\cite{Metsaev:2005ws}-\cite{Costa:2011mg}. In this paper we considered
the extended hamiltonian action for the bosonic totally symmetric fields.
Needless to say that generalization of our approach to the case of fermionic
fields \cite{Metsaev:2006zy} and mixed symmetry fields
\cite{Metsaev:1995re}-\cite{Buchbinder:2011xw} could also be of interest.

\bigskip

{\bf Acknowledgments}. This work was supported by the RFBR Grant
No.11-02-00814 and by the Alexander von Humboldt Foundation Grant PHYS0167.

%%%%%%%%%%%%%%%%%%%%%%%%%%%%%%%%%%%%%%%%%%%%%%%%%%%%%%%%%%%%%%%%%%%%%%%%%
\setcounter{section}{0}\setcounter{subsection}{0}
\appendix{ Notation }
%%%%%%%%%%%%%%%%%%%%%%%%%%%%%%%%%%%%%%%%%%%%%%%%%%%%%%%%%%%%%%%%%%%%%%%%%

Basis of $2\times 2$ matrices we use is defined as
\beq \label{manold-03112011-01}
&& \sigma_+  = \left(
\begin{array}{ll}
0 & 1
\\
0 & 0
\end{array}\right),
\quad
\sigma_-  = \left(
\begin{array}{ll}
0 & 0
\\
1 & 0
\end{array}\right),
\quad
\pi_+  = \left(
\begin{array}{ll}
1 & 0
\\
0 & 0
\end{array}\right),
\quad
\pi_-  = \left(
\begin{array}{ll}
0 & 0
\\
0 & 1
\end{array}\right)\,.
\eeq
Throughout the paper the notation $n \in [k]_2$
implies that $n =-k,-k+2,-k+4,\ldots,k-4, k-2,k$:
\be
\label{sumnot02} n \in [k]_2 \quad \Longrightarrow \quad n
=-k,-k+2,-k+4,\ldots,k-4, k-2,k\,.
\ee

{\bf Notation in basis of Lorentz algebra $so(d-1,1)$}. Our conventions are
as follows. $x^a$ denotes coordinates in $d$-dimensional flat space-time,
while $\partial_a$ denotes derivatives with respect to $x^a$, $\partial_a
\equiv
\partial /
\partial x^a$. Vector indices of the Lorentz algebra $so(d-1,1)$ take the
values $a,b,c,e=0,1,\ldots ,d-1$. We use the mostly positive flat metric
tensor $\eta^{ab}$. To simplify our expressions we drop $\eta_{ab}$ in scalar
products, i.e., we use $X^aY^a \equiv \eta_{ab}X^a Y^b$.

We use a set of the creation operators $\alpha^a$, $\alpha^z$,
$\zeta$, $\upsilon^\oplussm$, $\upsilon^\ominussm$ and the
respective set of annihilation operators $\bar{\alpha}^a$,
$\bar\alpha^z$, $\bar\zeta$, $\bar\upsilon^\ominussm$,
$\bar\upsilon^\oplussm$. These operators, to be
referred to as oscillators, satisfy the commutation relations%
\footnote{ Extensive study and applications of the oscillator formalism may
be fond in Refs.\cite{Bekaert:2006ix,Boulanger:2008up}.}
\beq
\label{man02-291102011-01}
&&{}\hspace{-1cm} [\bar\alpha^a,\alpha^b]=\eta^{ab}\,, \qquad
[\bar\alpha^z,\, \alpha^z]=1\,, \qquad [\bar\zeta,\zeta]=1\,,\qquad
[\bar\upsilon^\oplussm,\upsilon^\ominussm] =1\,,  \qquad
[\bar\upsilon^\ominussm,\upsilon^\oplussm] =1\,,  \quad
\\
\label{man02-291102011-02} && {}\hspace{-1cm} \bar\alpha^a |0\rangle
= 0\,,\qquad\quad \ \ \bar\alpha^z |0\rangle = 0\,,\qquad\quad \
\bar\zeta|0\rangle = 0\,, \qquad \ \bar\upsilon^\oplussm|0\rangle =
0\,, \qquad \quad \bar\upsilon^\ominussm|0\rangle =0\,.  \quad
\eeq
We adapt the following hermitian conjugation rules for the derivatives and
oscillators:
\be \label{03082011-01} \partial^{a\dagger} = - \partial^a\,, \qquad
\alpha^{a\dagger} = \bar\alpha^a\,, \qquad \alpha^{z\dagger} =
\bar\alpha^z\,, \qquad \zeta^\dagger = \bar\zeta \,,\qquad
\upsilon^{\oplussm\dagger} = \bar\upsilon^\oplussm\,,\qquad
\upsilon^{\ominussm\dagger} = \bar\upsilon^\ominussm\,.
\ee
We use operators constructed out of the derivatives and oscillators,
\beq
\label{manold-31102011-02}&& \Box = \partial^a\partial^a\,, \qquad\quad \ \
\alphabf\partialbf \equiv \alpha^a\partial^a\,,  \qquad
\bar\alphabf\partialbf \equiv \bar\alpha^a\partial^a\,,
\\
&& \alphabf^2 \equiv \alpha^a\alpha^a\,, \qquad\quad   \bar\alphabf^2 \equiv
\label{manold-31102011-02a1} \bar\alpha^a\bar\alpha^a\,,\qquad N_\alphabf
\equiv \alpha^a \bar\alpha^a \,,
\\
\label{manold-31102011-05} && N_z \equiv \alpha^z \bar\alpha^z \,,
\qquad \quad
N_\zeta \equiv \zeta \bar\zeta \,. \qquad\quad \ \ \
\eeq

{\bf Notation in basis of $so(d-1)$ algebra}. In the basis of $so(d-1)$
algebra, we split the space-time coordinates, derivatives, and oscillators as
follows
\beq
& x^a = t,x^i\,,\qquad \partial_a = \partial_t, \partial_i\,,\qquad
\partial_t\equiv \partial/\partial t\,,\qquad
\partial_i\equiv \partial/\partial x^i\,, &
\\
& \alpha^a = \alpha^0\,, \alpha^i\,,\qquad  \bar\alpha^a =
\bar\alpha^0\,,\bar\alpha^i\,,\qquad [\bar\alpha^0,\alpha^0] = -1\,, \qquad
[\bar\alpha^i,\alpha^j] =\delta^{ij}\,. &\eeq
Vector indices of the algebra $so(d-1)$ take the values $i,j=1,\ldots ,d-1$.
We use operators constructed out of the spatial derivative and oscillators,
\be
\alpha\partial \equiv \alpha^i\partial^i\,,\qquad\quad \bar\alpha\partial
\equiv \bar\alpha^i\partial^i\,,\qquad \alpha^2 \equiv \alpha^i\alpha^i\,,
\qquad\quad   \bar\alpha^2 \equiv \bar\alpha^i\bar\alpha^i\,,\qquad
N_\alpha\equiv\alpha^i\bar\alpha^i\,,
\ee

\beq
\label{man02-30112011-05} && n_{00} \equiv \sum_{n=0}^\infty
\frac{1}{(2n)!} \alpha^{2n}\alphab^{2n}\,,
\\
\label{man02-30112011-06} && n_{11} \equiv - \sum_{n=0}^\infty
\frac{2n+1}{(2n)!}\alpha^{2n}\alphab^{2n}\,,
\\
\label{man02-30112011-07} && n_{22} \equiv  \sum_{n=0}^\infty
\frac{2n+2}{(2n+1)!}\alpha^{2n}\alphab^{2n}\,,
\\
\label{man02-30112011-08} && n_{33} \equiv - \sum_{n=0}^\infty
\frac{4(n+1)^2}{(2n+3)!}\alpha^{2n}\alphab^{2n}\,.
\\
\label{man02-30112011-09} && n_{44} \equiv - \sum_{n=0}^\infty
\frac{1}{(2n+1)!}\alpha^{2n}\alphab^{2n}\,,
\\
\label{man02-30112011-10} && n_{02} \equiv
\alpha^2\sum_{n=0}^\infty
\frac{1}{(2n+1)!}\alpha^{2n}\alphab^{2n}\,,
\\
\label{man02-30112011-11} && n_{13} \equiv \alpha^2\sum_{n=0}^\infty
\frac{1}{(2n+1)!}\alpha^{2n}\alphab^{2n}\,,
\\
\label{man02-30112011-12} && \Ksf_0 \equiv \sum_{n=0}^\infty
\frac{1-2n }{(2n)!}\alpha^{2n}\alphab^{2n}\,,
\\
\label{man02-30112011-13} && \Ksf_3  \equiv \sum_{n=0}^\infty
\frac{2n+2 }{(2n+3)!}\alpha^{2n}\alphab^{2n}\,,
\eeq
\beq
\label{man02-30112011-14} && \Ksf_0\equiv n_{00} + \alpha^2 n_{44}
\alphab^2\,,
\\
\label{man02-30112011-15} &&  \Ksf_3 \equiv n_{33} - n_{44}\,,
\eeq
\beq
&& G_{01} =  \alpar -e_1 -\alpha^2\frac{1}{2N_\alpha+d-2}\eb_1\,,
\\
&& \Gb_{01} =  \albpar - \eb_1 - e_1 \frac{1}{2N_\alpha+d-2}\bar\alpha^2\,,
\\
&& G_{12} = \alpar - e_1 - \alpha^2 \frac{1}{2N_\alpha+d}\eb_1\,,
\\
&& \Gb_{12} = \albpar - \eb_1 - e_1 \frac{1}{2N_\alpha+d}\bar\alpha^2\,,
\\
&& G_{21}  = - \alpar + e_1 \frac{2N_\alpha+d}{2N_\alpha+d-2}\,,
\\
&& \Gb_{21}  = - \albpar + \frac{2N_\alpha+d}{2N_\alpha+d-2}\eb_1\,,
\\
&& G_{32} = 3\alpar + \alpha^2 \albpar  - e_1 (3 + \alpha^2
\frac{1}{2N_\alpha+d+4}\bar\alpha^2) - \alpha^2 \eb_1 \,,
\\
&& \Gb_{32} = 3\albpar + \alpar\bar\alpha^2  - (3 + \alpha^2
\frac{1}{2N_\alpha+d+4}\bar\alpha^2) \eb_1 - e_1\bar\alpha^2 \,,
\\
&&  G_{02}\equiv  -\alpha^2 n_{44} \Delta_\m - C_{10} n_{00}
C_{12}\,,
\\
&&  \Gb_{02}\equiv   n_{44}\bar\alpha^2 \Delta_\m + \Cb_{12} n_{00}
\Cb_{10}\,,
\\
&& G_{31}\equiv   -  G_{21} n_{44} C_{23} \,,
\\
&& \Gb_{31}\equiv   \Cb_{23} n_{44} \Gb_{21}\,,
\eeq
\beq
&& C_{10} = \alpar - e_1\,,
\\
&& \Cb_{10} = \albpar -\eb_1\,,
\\
&& C_{12} = \alpar -e_1 - \alpha^2 \frac{1}{2N_\alpha+d}\eb_1\,,
\\
&& \Cb_{12} = \albpar - \eb_1 -  e_1 \frac{1}{2N_\alpha+d}\bar\alpha^2\,,
\\
&& C_{21}  =  \alpar + \alpha^2\albpar  - (1 + \alpha^2
\frac{1}{2N_\alpha+d+2}\bar\alpha^2)e_1 - \alpha^2\eb_1 \,,
\\
&& \Cb_{21}  =  \albpar + \alpar\bar\alpha^2  - (1 + \alpha^2
\frac{1}{2N_\alpha+d+2}\bar\alpha^2)\eb_1 - e_1\bar\alpha^2\,,
\\
&& C_{23} =  \alpar - e_1 - \alpha^2 \frac{1}{2N_\alpha+d+2}\eb_1 \,,
\\
&& \Cb_{23} =  \albpar - \eb_1 -  e_1
\frac{1}{2N_\alpha+d+2}\bar\alpha^2 \,.
\eeq


\small
\begin{thebibliography}{30}

\parskip=-3pt


%\cite{Dirac:1951zz}
\bibitem{Dirac:1951zz}
  P.~A.~M.~Dirac,
  %``The Hamiltonian form of field dynamics,''
  Can.\ J.\ Math.\  {\bf 3}, 1 (1951).
  %%CITATION = CJMAA,3,1;%%


%\cite{Fradkin:1977hw}
\bibitem{Fradkin:1977hw}
  E.~S.~Fradkin and G.~A.~Vilkovisky,
  ``Quantization of relativistic systems with constraints: Equivalence of
  canonical and covariant formalisms in quantum theory of gravitational
  field,'', Preprint CERN-TH-2332, June 1977; Published in
  %%CITATION = CERN-TH-2332;%%
%\cite{Fradkin:2007zz}
%\bibitem{Fradkin:2007zz}
  E.~S.~Fradkin,
  ``Selected Papers On Theoretical Physics,''
%\href{http://www.slac.stanford.edu/spires/find/hep/www?irn=8542821}{SPIRES entry}
{\it  Moscow, Russia: Nauka (2007) 669 p}.




%\cite{Vasiliev:1990en}
\bibitem{Vasiliev:1990en}
  M.~A.~Vasiliev,
  %``Consistent equation for interacting gauge fields of all spins in
  %(3+1)-dimensions,''
  Phys.\ Lett.\  B {\bf 243}, 378 (1990);
  %%CITATION = PHLTA,B243,378;%%
%
%
%
%
%\cite{Vasiliev:2003ev}
%\bibitem{Vasiliev:2003ev}
%  M.~A.~Vasiliev,
  %``Nonlinear equations for symmetric massless higher spin fields in
  %(A)dS(d),''
  Phys.\ Lett.\  B {\bf 567}, 139 (2003)
  [arXiv:hep-th/0304049].
  %%CITATION = PHLTA,B567,139;%%


%\cite{Fronsdal:1978rb}
\bibitem{Fronsdal:1978rb}
  C.~Fronsdal,
  %``Massless Fields With Integer Spin,''
  Phys.\ Rev.\  D {\bf 18}, 3624 (1978).
  %%CITATION = PHRVA,D18,3624;%%



%\cite{Fronsdal:1978vb}
\bibitem{Fronsdal:1978vb}
C.~Fronsdal,
%``Singletons And Massless, Integral Spin Fields On De Sitter Space
%(Elementary Particles In A Curved Space Vii),''
Phys.\ Rev.\ D {\bf 20}, 848 (1979).
%%CITATION = PHRVA,D20,848;%%


%\cite{Aragone:1979hw}
\bibitem{Aragone:1979hw}
  C.~Aragone and S.~Deser,
  %``Hamiltonian Form For Massless Higher Spin Fermions,''
  Phys.\ Rev.\  D {\bf 21}, 352 (1980);
  %%CITATION = PHRVA,D21,352;%%
%
%\cite{Aragone:1980rk}
%\bibitem{Aragone:1980rk}
%  C.~Aragone and S.~Deser,
  %``Higher Spin Vierbein Gauge Fermions And Hypergravities,''
  Nucl.\ Phys.\  B {\bf 170}, 329 (1980).
  %%CITATION = NUPHA,B170,329;%%



%\cite{Vasiliev:1987hv}
\bibitem{Vasiliev:1987hv}
  M.~A.~Vasiliev,
  %``LINEARIZED CURVATURES FOR AUXILIARY FIELDS IN THE DE SITTER SPACE,''
  Nucl.\ Phys.\  B {\bf 307}, 319 (1988).
  %%CITATION = NUPHA,B307,319;%%


%\cite{Rindani:1988wf}
\bibitem{Rindani:1988wf}
  S.~D.~Rindani and M.~Sivakumar,
  %``HAMILTONIAN FORMULATION OF A GAUGE INVARIANT MASSIVE SPIN 3/2 THEORY,''
  Phys.\ Rev.\  D {\bf 37}, 3543 (1988).
  %%CITATION = PHRVA,D37,3543;%%


%\cite{Lee:1982cp}
\bibitem{Lee:1982cp}
  S.~C.~Lee and P.~van Nieuwenhuizen,
  %``Counting Of States In Higher Derivative Field Theories,''
  Phys.\ Rev.\  D {\bf 26}, 934 (1982).
  %%CITATION = PHRVA,D26,934;%%




%\cite{Metsaev:2008fs}
\bibitem{Metsaev:2008fs}
  R.~R.~Metsaev,
  %``Shadows, currents and AdS,''
  Phys.\ Rev.\  D {\bf 78}, 106010 (2008)
  [arXiv:0805.3472 [hep-th]].
  %%CITATION = PHRVA,D78,106010;%%



%\cite{Metsaev:2008ks}
\bibitem{Metsaev:2008ks}
  R.R. Metsaev,
  %``CFT adapted gauge invariant formulation of arbitrary spin fields in AdS and
  %modified de Donder gauge,''
  Phys. Lett. B {\bf 671}, 128 (2009)
  [arXiv:0808.3945].
  %%CITATION = PHLTA,B671,128;%%


%\cite{Metsaev:2009hp}
\bibitem{Metsaev:2009hp}
  R.R. Metsaev,
  %``CFT adapted gauge invariant formulation of massive arbitrary spin fields in
  %AdS,''
  Phys. Lett. B {\bf 682}, 455 (2010)
  [arXiv:0907.2207].
  %%CITATION = PHLTA,B682,455;%%


%\cite{Metsaev:2007fq}
\bibitem{Metsaev:2007fq}
  R.~R.~Metsaev,
  %``Ordinary-derivative formulation of conformal low spin fields,''
  JHEP {\bf 1201}, 064 (2012)
  [arXiv:0707.4437 [hep-th]].
  %%CITATION = ARXIV:0707.4437;%%

%\cite{Metsaev:2007rw}
\bibitem{Metsaev:2007rw}
  R.~R.~Metsaev,
  %``Ordinary-derivative formulation of conformal
  %totally symmetric arbitrary spin bosonic fields,''
  JHEP {\bf 1206}, 062 (2012)
  [arXiv:0709.4392 [hep-th]].
  %%CITATION = ARXIV:0709.4392;%%




%\cite{Zinoviev:2001dt}
\bibitem{Zinoviev:2001dt}
  Yu.~M.~Zinoviev,
  ``On massive high spin particles in (A)dS,''
  arXiv:hep-th/0108192.
  %%CITATION = HEP-TH 0108192;%%




%\cite{Lopatin:1987hz}
\bibitem{Lopatin:1987hz}
  V.~E.~Lopatin and M.~A.~Vasiliev,
  %``FREE MASSLESS BOSONIC FIELDS OF ARBITRARY SPIN IN d-DIMENSIONAL DE SITTER
  %SPACE,''
  Mod.\ Phys.\ Lett.\  A {\bf 3}, 257 (1988).
  %%CITATION = MPLAE,A3,257;%%


%\cite{Metsaev:1999ui}
\bibitem{Metsaev:1999ui}
  R.~R.~Metsaev,
  %``Light cone form of field dynamics in anti-de Sitter spacetime and  AdS/CFT
  %correspondence,''
  Nucl.\ Phys.\ B {\bf 563}, 295 (1999)
  hep-th/9906217.
  %%CITATION = HEP-TH 9906217;%%



%\cite{Buchbinder:2001bs}
\bibitem{Buchbinder:2001bs}
  I.~L.~Buchbinder, A.~Pashnev and M.~Tsulaia,
  %``Lagrangian formulation of the massless higher integer spin fields in  the
  %AdS background,''
  Phys.\ Lett.\  B {\bf 523}, 338 (2001)
  [arXiv:hep-th/0109067].
  %%CITATION = PHLTA,B523,338;%%




%\cite{Rindani:1985pi}
\bibitem{Rindani:1985pi}
  S.~D.~Rindani and M.~Sivakumar,
  %``Gauge - Invariant Description Of Massive Higher - Spin Particles By
  %Dimensional Reduction,''
  Phys.\ Rev.\  D {\bf 32}, 3238 (1985).
  %%CITATION = PHRVA,D32,3238;%%



%\cite{Aragone:1988yx}
\bibitem{Aragone:1988yx}
  C.~Aragone, S.~Deser and Z.~Yang,
  %``MASSIVE HIGHER SPIN FROM DIMENSIONAL REDUCTION OF GAUGE FIELDS,''
  Annals Phys.\  {\bf 179}, 76 (1987).
  %%CITATION = APNYA,179,76;%%


%\cite{Rindani:1988gb}
\bibitem{Rindani:1988gb}
  S.~D.~Rindani, D.~Sahdev and M.~Sivakumar,
  %``Dimensional reduction of symmetric higher spin actions. 1. Bosons,''
  Mod.\ Phys.\ Lett.\  A {\bf 4}, 265 (1989).
  %%CITATION = MPLAE,A4,265;%%



%\cite{Metsaev:2000qb}
\bibitem{Metsaev:2000qb}
  R.~R.~Metsaev,
  %``Massive fields in AdS(3) and compactification in AdS spacetime,''
  Nucl.\ Phys.\ Proc.\ Suppl.\  {\bf 102}, 100 (2001)
  [arXiv:hep-th/0103088].
  %%CITATION = NUPHZ,102,100;%%


%\cite{Biswas:2002nk}
\bibitem{Biswas:2002nk}
  T.~Biswas and W.~Siegel,
  %``Radial dimensional reduction: (Anti) de Sitter theories from flat,''
  JHEP {\bf 0207}, 005 (2002)
  [arXiv:hep-th/0203115].
  %%CITATION = JHEPA,0207,005;%%



%\cite{Artsukevich:2008vy}
\bibitem{Artsukevich:2008vy}
  A.~Y.~Artsukevich and M.~A.~Vasiliev,
  %``On Dimensional Degression in AdS(d),''
  Phys.\ Rev.\  D {\bf 79}, 045007 (2009)
  [arXiv:0810.2065 [hep-th]].
  %%CITATION = PHRVA,D79,045007;%%


%\cite{Buchbinder:2005ua}
\bibitem{Buchbinder:2005ua}
  I.~L.~Buchbinder and V.~A.~Krykhtin,
  %``Gauge invariant Lagrangian construction for massive bosonic higher spin
  %fields in D dimensions,''
  Nucl.\ Phys.\ B {\bf 727}, 537 (2005)
  [arXiv:hep-th/0505092].
  %%CITATION = HEP-TH 0505092;%%



%\cite{Buchbinder:2006ge}
\bibitem{Buchbinder:2006ge}
  I.~L.~Buchbinder, V.~A.~Krykhtin and P.~M.~Lavrov,
  %``\hfill{\normalsize{}hep-th/0608005,''
  Nucl.\ Phys.\  B {\bf 762}, 344 (2007)
  hep-th/0608005
  %%CITATION = NUPHA,B762,344;%%



%\cite{Alkalaev:2009vm}
\bibitem{Alkalaev:2009vm}
  K.~B.~Alkalaev and M.~Grigoriev,
  %``Unified BRST description of AdS gauge fields,''
  Nucl.\ Phys.\  B {\bf 835}, 197 (2010)
  [arXiv:0910.2690 [hep-th]].
  %%CITATION = NUPHA,B835,197;%%
%
%\cite{Alkalaev:2011zv}
%\bibitem{Alkalaev:2011zv}
% K.~Alkalaev and M.~Grigoriev,
  %``Unified BRST approach to (partially) massless and massive AdS fields of
  %arbitrary symmetry type,''
  Nucl.\ Phys.\  B {\bf 853}, 663 (2011)
  [arXiv:1105.6111 [hep-th]].
  %%CITATION = NUPHA,B853,663;%%


%\cite{Grigoriev:2011gp}
\bibitem{Grigoriev:2011gp}
  M.~Grigoriev and A.~Waldron,
  %``Massive Higher Spins from BRST and Tractors,''
  Nucl.\ Phys.\  B {\bf 853}, 291 (2011)
  [arXiv:1104.4994 [hep-th]].
  %%CITATION = NUPHA,B853,291;%%



%\cite{Zinoviev:2008ze}
\bibitem{Zinoviev:2008ze}
  Yu.~M.~Zinoviev,
  %``Frame-like gauge invariant formulation for massive high spin particles,''
  Nucl.\ Phys.\  B {\bf 808}, 185 (2009)
  [arXiv:0808.1778 [hep-th]].
  %%CITATION = NUPHA,B808,185;%%


%\cite{Ponomarev:2010st}
\bibitem{Ponomarev:2010st}
  D.~S.~Ponomarev and M.~A.~Vasiliev,
  %``Frame-Like Action and Unfolded Formulation for Massive Higher-Spin
  %Fields,''
  Nucl.\ Phys.\  B {\bf 839}, 466 (2010)
  [arXiv:1001.0062 [hep-th]].
  %%CITATION = NUPHA,B839,466;%%



%\cite{Metsaev:2002vr}
\bibitem{Metsaev:2002vr}
  R.~R.~Metsaev,
  %``Massless arbitrary spin fields in AdS(5),''
  Phys.\ Lett.\  {\bf B531}, 152-160 (2002).
  [hep-th/0201226].


%\cite{Metsaev:2003cu}
\bibitem{Metsaev:2003cu}
  R.~R.~Metsaev,
  %``Massive totally symmetric fields in AdS(d),''
  Phys. Lett.  B {\bf 590}, 95 (2004)
  hep-th/0312297
  %%CITATION = PHLTA,B590,95;%%

%\cite{Metsaev:2004ee}
\bibitem{Metsaev:2004ee}
  R.~R.~Metsaev,
  %``Mixed symmetry massive fields in AdS(5),''
  Class.\ Quant.\ Grav.\  {\bf 22}, 2777 (2005)
  [arXiv:hep-th/0412311].
  %%CITATION = CQGRD,22,2777;%%


%\cite{Gutsche:2011bu}
\bibitem{Gutsche:2011bu}
  T.~Gutsche, V.~E.~Lyubovitskij, I.~Schmidt and A.~Vega,
  %``Mesons and baryons in a soft-wall holographic approach,''
  arXiv:1108.0527;
  %%CITATION = ARXIV:1108.0527;%%
%
%\cite{Gutsche:2011vb}
%\bibitem{Gutsche:2011vb}
%  T.~Gutsche, V.~E.~Lyubovitskij, I.~Schmidt and A.~Vega,
  %``Dilaton in a soft-wall holographic approach to mesons and baryons,''
  arXiv:1108.0346.
  %%CITATION = ARXIV:1108.0346;%%


%\cite{Guttenberg:2008qe}
\bibitem{Guttenberg:2008qe}
S.~Guttenberg and G.~Savvidy,
% ``Schwinger-Fronsdal Theory of Abelian Tensor Gauge Fields,''
SIGMAP bulletin 4, 061 (2008) arXiv:0804.0522 [hep-th].
  %%CITATION = ARXIV:0804.0522;%%


%\cite{Manvelyan:2008ks}
\bibitem{Manvelyan:2008ks}
  R.~Manvelyan, K.~Mkrtchyan and W.~Ruhl,
  %``Ultraviolet behaviour of higher spin gauge field propagators and one loop
  %mass renormalization,''
  Nucl.\ Phys.\  B {\bf 803}, 405 (2008)
  [arXiv:0804.1211 [hep-th]].
  %%CITATION = NUPHA,B803,405;%%




%\cite{Fotopoulos:2009iw}
\bibitem{Fotopoulos:2009iw}
  A.~Fotopoulos and M.~Tsulaia,
  %``Current Exchanges for Reducible Higher Spin Multiplets and Gauge Fixing,''
  JHEP {\bf 0910}, 050 (2009)
  [arXiv:0907.4061 [hep-th]].
  %%CITATION = JHEPA,0910,050;%%



%\cite{Metsaev:2010kp}
\bibitem{Metsaev:2010kp}
  R.~R.~Metsaev,
  %``6d conformal gravity,''
  J.\ Phys.\ A  {\bf 44}, 175402 (2011)
  [arXiv:1012.2079 [hep-th]].
  %%CITATION = JPAGB,A44,175402;%%



%\cite{Metsaev:2009ym}
\bibitem{Metsaev:2009ym}
  R.~R.~Metsaev,
  %``Gauge invariant two-point vertices of shadow fields, AdS/CFT, and conformal
  %fields,''
  Phys.\ Rev.\  D {\bf 81}, 106002 (2010)
  [arXiv:0907.4678 [hep-th]].
  %%CITATION = PHRVA,D81,106002;%%


%\cite{Metsaev:2010zu}
\bibitem{Metsaev:2010zu}
  R.~R.~Metsaev,
  %``Gauge invariant approach to low-spin anomalous conformal currents and
  %shadow fields,''
  Phys.\ Rev.\  D {\bf 83}, 106004 (2011)
  [arXiv:1011.4261 [hep-th]].
  %%CITATION = PHRVA,D83,106004;%%

%\cite{Metsaev:2011uy}
\bibitem{Metsaev:2011uy}
  R.~R.~Metsaev,
  %``Anomalous conformal currents, shadow fields and massive AdS fields,''
  Phys.\ Rev.\ D {\bf 85}, 126011 (2012)
  [arXiv:1110.3749 [hep-th]].
  %%CITATION = ARXIV:1110.3749;%%



%\cite{Bekaert:2009fg}
\bibitem{Bekaert:2009fg}
  X.~Bekaert and M.~Grigoriev,
  %``Manifestly Conformal Descriptions and Higher Symmetries of Bosonic
  %Singletons,''
  SIGMA {\bf 6}, 038 (2010)
  [arXiv:0907.3195 [hep-th]].
  %%CITATION = 00480,6,038;%%


%\cite{Bonezzi:2010jr}
\bibitem{Bonezzi:2010jr}
  R.~Bonezzi, E.~Latini and A.~Waldron,
  %``Gravity, Two Times, Tractors, Weyl Invariance and Six Dimensional Quantum
  %Mechanics,''
  Phys.\ Rev.\  D {\bf 82}, 064037 (2010)
  [arXiv:1007.1724 [hep-th]].
  %%CITATION = PHRVA,D82,064037;%%


%\cite{Bekaert:2011js}
\bibitem{Bekaert:2011js}
  X.~Bekaert,
  %``Singletons and their maximal symmetry algebras,''
  arXiv:1111.4554 [math-ph].
  %%CITATION = ARXIV:1111.4554;%%


%\cite{Fotopoulos:2006ci}
\bibitem{Fotopoulos:2006ci}
  A.~Fotopoulos, K.~L.~Panigrahi and M.~Tsulaia,
  %``Lagrangian Formulation Of Higher Spin Theories On Ads Space,''
  Phys.\ Rev.\  D {\bf 74}, 085029 (2006)
  [hep-th/0607248].
  %%CITATION = PHRVA,D74,085029;%%


%\cite{Francia:2002aa}
\bibitem{Francia:2002aa}
  D.~Francia and A.~Sagnotti,
  %``Free geometric equations for higher spins,''
  Phys.\ Lett.\ B {\bf 543}, 303 (2002)
  [arXiv:hep-th/0207002].
  %%CITATION = HEP-TH 0207002;%%



%\cite{Sagnotti:2003qa}
\bibitem{Sagnotti:2003qa}
  A.~Sagnotti and M.~Tsulaia,
  %``On higher spins and the tensionless limit of string theory,''
  Nucl.\ Phys.\  B {\bf 682}, 83 (2004)
  [arXiv:hep-th/0311257].
  %%CITATION = NUPHA,B682,83;%%





%\cite{Buchbinder:2007ak}
\bibitem{Buchbinder:2007ak}
  I.L.~Buchbinder, A.V.~Galajinsky and V.~A.~Krykhtin,
  %``Quartet unconstrained formulation for massless higher spin fields,''
  Nucl.Phys. B{\bf 779}, 155 (2007)
  [hep-th/0702161].
  %%CITATION = NUPHA,B779,155;%%

%\cite{Buchbinder:2008ss}
\bibitem{Buchbinder:2008ss}
  I.~L.~Buchbinder and A.~V.~Galajinsky,
  %``Quartet unconstrained formulation for massive higher spin fields,''
  JHEP {\bf 0811}, 081 (2008)
  [arXiv:0810.2852 [hep-th]].
  %%CITATION = JHEPA,0811,081;%%



%\cite{Campoleoni:2008jq}
\bibitem{Campoleoni:2008jq}
  A.Campoleoni, D.Francia, J.Mourad and A.Sagnotti,
  %``Unconstrained Higher Spins of Mixed Symmetry. I. Bose Fields,''
  Nucl.Phys.B {\bf 815}, 289 (2009)
  [arXiv:0810.4350].
  %%CITATION = NUPHA,B815,289;%%


%\cite{Bolotin:1999fa}
\bibitem{Bolotin:1999fa}
  K.~I.~Bolotin and M.~A.~Vasiliev,
  %``Star-product and massless free field dynamics in AdS(4),''
  Phys.\ Lett.\  B {\bf 479}, 421 (2000)
  [arXiv:hep-th/0001031].
  %%CITATION = PHLTA,B479,421;%%



%\cite{Didenko:2009td}
\bibitem{Didenko:2009td}
  V.~E.~Didenko and M.~A.~Vasiliev,
  %``Static BPS black hole in 4d higher-spin gauge theory,''
  Phys.\ Lett.\  B {\bf 682}, 305 (2009)
  [arXiv:0906.3898 [hep-th]].
  %%CITATION = PHLTA,B682,305;%%


%\cite{Didenko:2011ir}
\bibitem{Didenko:2011ir}
  V.~E.~Didenko,
  %``Coordinate independent approach to 5d black holes,''
  Class.\ Quant.\ Grav.\  {\bf 29}, 025009 (2012)
  [arXiv:1108.4321 [hep-th]].
  %%CITATION = ARXIV:1108.4321;%%



%\cite{Fradkin:1987ks}
\bibitem{Fradkin:1987ks}
  E.~S.~Fradkin and M.~A.~Vasiliev,
  %``On the Gravitational Interaction of Massless Higher Spin Fields,''
  Phys.\ Lett.\  B {\bf 189}, 89 (1987).
  %%CITATION = PHLTA,B189,89;%%



%\cite{Metsaev:2005ar}
\bibitem{Metsaev:2005ar}
  R.~R.~Metsaev,
  %``Cubic interaction vertices for massive and massless higher spin fields,''
  Nucl.\ Phys.\  B {\bf 759}, 147 (2006)
  [arXiv:hep-th/0512342];
  %%CITATION = NUPHA,B759,147;%%


%\cite{Metsaev:1993ap}
\bibitem{Metsaev:1993ap}
  R.~R.~Metsaev,
  %``Generating function for cubic interaction vertices of higher spin fields in
  %any dimension,''
  Mod.\ Phys.\ Lett.\  A {\bf 8}, 2413 (1993).
  %%CITATION = MPLAE,A8,2413;%%


%\cite{Metsaev:2004wv}
\bibitem{Metsaev:2004wv}
  R.~R.~Metsaev,
  %``Eleven dimensional supergravity in light cone gauge,''
  Phys.\ Rev.\  D {\bf 71}, 085017 (2005)
  [arXiv:hep-th/0410239].
  %%CITATION = PHRVA,D71,085017;%%




%\cite{Metsaev:2006ui}
\bibitem{Metsaev:2006ui}
  R.~R.~Metsaev,
  %``Gravitational and higher-derivative interactions of massive spin 5/2 field
  %in (A)dS space,''
  Phys.\ Rev.\  D {\bf 77}, 025032 (2008)
  [arXiv:hep-th/0612279].
  %%CITATION = PHRVA,D77,025032;%%

%\cite{Polyakov:2010qs}
\bibitem{Polyakov:2010qs}
  D.~Polyakov,
  %``Gravitational Couplings of Higher Spins from String Theory,''
  Int.\ J.\ Mod.\ Phys.\  A {\bf 25}, 4623 (2010);
  [arXiv:1005.5512 [hep-th]].
  %%CITATION = IMPAE,A25,4623;%%
%
%\cite{Polyakov:2010sk}
%\bibitem{Polyakov:2010sk}
%  D.~Polyakov,
  %``Higher Spins and Open Strings: Quartic Interactions,''
  Phys.\ Rev.\  D {\bf 83}, 046005 (2011)
  [arXiv:1011.0353 [hep-th]].
  %%CITATION = PHRVA,D83,046005;%%


%\cite{Bekaert:2010hp}
\bibitem{Bekaert:2010hp}
  X.~Bekaert, N.~Boulanger and S.~Leclercq,
  %``Strong obstruction of the Berends-Burgers-van Dam spin-3 vertex,''
  J.\ Phys.\ A  {\bf 43}, 185401 (2010)
  [arXiv:1002.0289 [hep-th]].
  %%CITATION = JPAGB,A43,185401;%%




%\cite{Sagnotti:2010at}
\bibitem{Sagnotti:2010at}
  A.~Sagnotti and M.~Taronna,
  %``String Lessons for Higher-Spin Interactions,''
  Nucl.\ Phys.\  B {\bf 842}, 299 (2011)
  [arXiv:1006.5242 [hep-th]].
  %%CITATION = NUPHA,B842,299;%%



%\cite{Manvelyan:2010je}
\bibitem{Manvelyan:2010je}
  R.~Manvelyan, K.~Mkrtchyan and W.~Ruehl,
  %``A generating function for the cubic interactions of higher spin fields,''
  Phys.\ Lett.\  B {\bf 696}, 410 (2011)
  [arXiv:1009.1054 [hep-th]].
  %%CITATION = PHLTA,B696,410;%%

%\cite{Alkalaev:2010af}
\bibitem{Alkalaev:2010af}
  K.~Alkalaev,
  %``FV-type action for AdS(5) mixed-symmetry fields,''
  JHEP {\bf 1103}, 031 (2011)
  [arXiv:1011.6109 [hep-th]].
  %%CITATION = JHEPA,1103,031;%%


%\cite{Boulanger:2011se}
\bibitem{Boulanger:2011se}
  N.~Boulanger, E.~D.~Skvortsov,
  %``Higher-spin algebras and cubic interactions
  %for simple mixed-symmetry fields in AdS spacetime,''
  JHEP {\bf 1109}, 063 (2011).
  [arXiv:1107.5028 [hep-th]].

%\cite{Boulanger:2011qt}
\bibitem{Boulanger:2011qt}
  N.~Boulanger, E.D.~Skvortsov, Y.M.~Zinoviev,
  %``Gravitational cubic interactions for a simple
  %mixed-symmetry gauge field in AdS and flat backgrounds,''
  J.Phys. A {\bf A44}, 415403 (2011).
  [1107.1872 [hep-th]].


%\cite{Vasilev:2011xf}
\bibitem{Vasilev:2011xf}
  M.~A.~Vasiliev,
  %``Cubic Vertices for Symmetric Higher-Spin Gauge Fields in $(A)dS_d$,''
  Nucl.\ Phys.\ B {\bf 862}, 341 (2012)
  [arXiv:1108.5921 [hep-th]].
  %%CITATION = ARXIV:1108.5921;%%




%\cite{Ruehl:2011tk}
\bibitem{Ruehl:2011tk}
  W.~Ruehl,
  %``Solving Noether's equations for gauge invariant local Lagrangians of N
  %arbitrary higher even spin fields,''
  arXiv:1108.0225 [hep-th].
  %%CITATION = ARXIV:1108.0225;%%


%\cite{Joung:2011ww}
\bibitem{Joung:2011ww}
  E.~Joung and M.~Taronna,
  %``Cubic interactions of massless higher spins in (A)dS: metric-like approach,''
  Nucl.\ Phys.\ B {\bf 861}, 145 (2012)
  [arXiv:1110.5918 [hep-th]].
  %%CITATION = ARXIV:1110.5918;%%



%\cite{Metsaev:1998it}
\bibitem{Metsaev:1998it}
  R.~R.~Metsaev and A.~A.~Tseytlin,
  %``Type IIB superstring action in AdS(5) x S(5) background,''
  Nucl.\ Phys.\  B {\bf 533}, 109 (1998)
  [arXiv:hep-th/9805028].
  %%CITATION = NUPHA,B533,109;%%




%\cite{Metsaev:2000yf}
\bibitem{Metsaev:2000yf}
  R.~R.~Metsaev and A.~A.~Tseytlin,
  %``Superstring action in AdS(5) x S(5): kappa-symmetry light cone gauge,''
  Phys.\ Rev.\  D {\bf 63}, 046002 (2001)
  [arXiv:hep-th/0007036].
  %%CITATION = PHRVA,D63,046002;%%



%\cite{Metsaev:2000mv}
\bibitem{Metsaev:2000mv}
  R.~R.~Metsaev and A.~A.~Tseytlin,
  %``Superparticle and superstring in AdS(3) x S(3) Ramond-Ramond  background in
  %light-cone gauge,''
  J.\ Math.\ Phys.\  {\bf 42}, 2987 (2001)
  [arXiv:hep-th/0011191].
  %%CITATION = JMAPA,42,2987;%%




%\cite{Metsaev:2005ws}
\bibitem{Metsaev:2005ws}
  R.~R.~Metsaev,
  %``Light-cone formulation of conformal field theory adapted to AdS/CFT
  %correspondence,''
  Phys.\ Lett.\  B {\bf 636}, 227 (2006)
  [arXiv:hep-th/0512330].
  %%CITATION = PHLTA,B636,227;%%


%\cite{Chang:2011mz}
\bibitem{Chang:2011mz}
  C.~-M.~Chang, X.~Yin,
  %``Higher Spin Gravity with Matter in AdS_3 and Its CFT Dual,''
  [arXiv:1106.2580 [hep-th]].


%\cite{Koch:2010cy}
\bibitem{Koch:2010cy}
  R.d.M.Koch, A.Jevicki, K.Jin and J.P.Rodrigues,
  %``$AdS_4/CFT_3$ Construction from Collective Fields,''
  Phys.Rev. D {\bf 83}, 025006 (2011)
  [arXiv:1008.0633].
  %%CITATION = PHRVA,D83,025006;%%


%\cite{Costa:2011mg}
\bibitem{Costa:2011mg}
  M.~S.~Costa, J.~Penedones, D.~Poland and S.~Rychkov,
  %``Spinning Conformal Correlators,''
  arXiv:1107.3554 [hep-th];
  %%CITATION = ARXIV:1107.3554;%%
%
%\cite{Costa:2011dw}
%\bibitem{Costa:2011dw}
%  M.~S.~Costa, J.~Penedones, D.~Poland and S.~Rychkov,
  %``Spinning Conformal Blocks,''
  arXiv:1109.6321.
  %%CITATION = ARXIV:1109.6321;%%






%\cite{Metsaev:2006zy}
\bibitem{Metsaev:2006zy}
  R.~R.~Metsaev,
  %``Gauge invariant formulation of massive totally symmetric fermionic fields
  %in (A)dS space,''
  Phys.\ Lett.\  B {\bf 643}, 205 (2006)
  [arXiv:hep-th/0609029].
  %%CITATION = PHLTA,B643,205;%%








%\cite{Metsaev:1995re}
\bibitem{Metsaev:1995re}
  R.~R.~Metsaev,
  %``Massless mixed symmetry bosonic free fields
  %in d-dimensional anti-de Sitter
  %space-time,''
  Phys.\ Lett.\ B {\bf 354}, 78 (1995).
  %%CITATION = PHLTA,B354,78;%%


%\cite{Metsaev:1995jp}
\bibitem{Metsaev:1995jp}
  R.~R.~Metsaev,
  %``All Conformal Invariant Representations Of D-Dimensional Anti-De Sitter
  %Group,''
  Mod.\ Phys.\ Lett.\  A {\bf 10}, 1719 (1995).
  %%CITATION = MPLAE,A10,1719;%%



%\cite{Alkalaev:2003qv}
\bibitem{Alkalaev:2003qv}
K.~B.~Alkalaev, O.~V.~Shaynkman and M.~A.~Vasiliev,
%``On the frame-like formulation of mixed-symmetry massless fields in
%(A)dS(d),''
Nucl.\ Phys.\  B {\bf 692}, 363 (2004) [arXiv:hep-th/0311164];
%%CITATION = NUPHA,B692,363;%%
%
%
%\cite{Alkalaev:2006rw}
%\bibitem{Alkalaev:2006rw}
%K.~B.~Alkalaev, O.~V.~Shaynkman and M.~A.~Vasiliev,
%``Frame-like formulation for free mixed-symmetry bosonic massless
%higher-spin fields in AdS(d),''
arXiv:hep-th/0601225.
%%CITATION = HEP-TH/0601225;%%






%\cite{Zinoviev:2008ve}
\bibitem{Zinoviev:2008ve}
  Yu.~M.~Zinoviev,
  %``Toward frame-like gauge invariant formulation for massive mixed symmetry
  %bosonic fields,''
  Nucl.\ Phys.\  B {\bf 812}, 46 (2009)
  [arXiv:0809.3287 [hep-th]].
  %%CITATION = NUPHA,B812,46;%%


%\cite{Skvortsov:2008vs}
\bibitem{Skvortsov:2008vs}
  E.~D.~Skvortsov,
  %``Mixed-Symmetry Massless Fields in Minkowski space Unfolded,''
  JHEP {\bf 0807}, 004 (2008)
  [arXiv:0801.2268 [hep-th]].
  %%CITATION = JHEPA,0807,004;%%
%
%\cite{Skvortsov:2008sh}
%\bibitem{Skvortsov:2008sh}
%  E.~D.~Skvortsov,
  %``Frame-like Actions for Massless Mixed-Symmetry Fields in Minkowski space,''
  Nucl.\ Phys.\  B {\bf 808}, 569 (2009)
  [arXiv:0807.0903 [hep-th]].
  %%CITATION = NUPHA,B808,569;%%
%
%\cite{Skvortsov:2009nv}
%\bibitem{Skvortsov:2009nv}
%  E.~D.~Skvortsov,
  %``Gauge fields in (A)dS within the unfolded approach: algebraic aspects,''
  JHEP {\bf 1001}, 106 (2010)
  [arXiv:0910.3334 [hep-th]].
  %%CITATION = JHEPA,1001,106;%%


%\cite{Buchbinder:2011xw}
\bibitem{Buchbinder:2011xw}
  I.~L.~Buchbinder and A.~Reshetnyak,
  %``General Lagrangian Formulation for Higher
  %Spin Fields with Arbitrary Index Symmetry. I. Bosonic fields,''
  Nucl.\ Phys.\ B {\bf 862}, 270 (2012)
  [arXiv:1110.5044 [hep-th]].
  %%CITATION = ARXIV:1110.5044;%%





%\cite{Bekaert:2006ix}
\bibitem{Bekaert:2006ix}
  X.~Bekaert and N.~Boulanger,
  %``Tensor gauge fields in arbitrary representations of GL(D,R). II. Quadratic
  %actions,''
  Commun.\ Math.\ Phys.\  {\bf 271}, 723 (2007)
  [arXiv:hep-th/0606198].
  %%CITATION = CMPHA,271,723;%%

%\cite{Boulanger:2008up}
\bibitem{Boulanger:2008up}
  N.~Boulanger, C.~Iazeolla and P.~Sundell,
  %``Unfolding Mixed-Symmetry Fields in AdS and the BMV Conjecture: I. General
  %Formalism,''
  JHEP {\bf 0907}, 013 (2009)
  [arXiv:0812.3615 [hep-th]].
  %%CITATION = JHEPA,0907,013;%%
%
%\cite{Boulanger:2008kw}
%\bibitem{Boulanger:2008kw}
%  N.~Boulanger, C.~Iazeolla and P.~Sundell,
  %``Unfolding Mixed-Symmetry Fields in AdS and the BMV Conjecture: II.
  %Oscillator Realization,''
  JHEP {\bf 0907}, 014 (2009)
  [arXiv:0812.4438 [hep-th]].
  %%CITATION = JHEPA,0907,014;%%









\end{thebibliography}
\end{document}